\documentclass[twocolumn,aps,prx,showpacs,superscriptaddress,letterpaper]{revtex4-1}  
\usepackage[applemac]{inputenc}
\usepackage[english]{babel}
\usepackage{graphicx}
\usepackage{epstopdf}
\usepackage{color}
\usepackage{soul}
\usepackage{amsmath,amsfonts, amssymb} 
\usepackage{fancyhdr}
\usepackage{enumerate}
\usepackage{hyperref}

\definecolor{rltred}{rgb}{0.75,0,0}
\definecolor{rosepale}{rgb}{1,.8,.8}
\definecolor{rltgreen}{rgb}{0,0.5,0}
\definecolor{oneblue}{rgb}{0,0,0.75}
\definecolor{marron}{rgb}{0.64,0.16,0.16}
\definecolor{forestgreen}{rgb}{0.13,0.54,0.13}
\definecolor{purple}{rgb}{0.62,0.12,0.94}
\definecolor{dockerblue}{rgb}{0.11,0.56,0.98}
\definecolor{freeblue}{rgb}{0.25,0.41,0.88}
\definecolor{myblue}{rgb}{0,0.2,0.4}
\definecolor{grey}{rgb}{0.25,0.25,0.25}
\definecolor{ww}{rgb}{1,1,1}

\sethlcolor{yellow}
\newcommand{\SiN}{{Si$_3$N$_4$}}
\newcommand{\omegac}{\omega_c}
\newcommand{\gsq}{g^{\prime}}
\newcommand{\gsqzpf}{\tilde{g}^{\prime}}
\newcommand{\glin}{g}
\newcommand{\glinzpf}{\tilde{\glin}}
\newcommand{\xzpf}{x_{\text{zpf}}}
\newcommand{\meff}{m}
\newcommand{\omegam}{\omega_{m}}
\newcommand{\Qm}{Q_{m}}
\newcommand{\omegamnot}{\omega_{m}}
\newcommand{\alphacap}{\alpha_{\text{cap}}}

\newcommand{\Hnot}{\hat{H}_{0}}
\newcommand{\Hint}{\hat{H}_{J}}
\newcommand{\HOM}{\hat{H}_{\text{OM}}}
\newcommand{\Pin}{P_{\text{in}}}
\newcommand{\Gammath}{\Gamma_{\text{th}}}
\newcommand{\Gammaopt}{\Gamma_{\text{opt}}}
\newcommand{\Gammameas}{\Gamma_{\text{meas}}}
\newcommand{\Tbath}{T_{b}}
\newcommand{\nbath}{\bar{n}_{\text{th}}}
\newcommand{\gammami}{\gamma_{i}}

\DeclareMathOperator{\sgn}{sgn}
\DeclareMathOperator{\SNR}{SNR}

\usepackage[none]{hyphenat}

\begin{document}

\title{Position-squared coupling in a tunable photonic crystal optomechanical cavity}

\author{Taofiq K. Para\"iso} 
\affiliation{Max Planck Institute for the Science of Light, G\"unther-Scharowsky-Stra\ss e 1/Bau 24, D-91058 Erlangen, Germany}
\affiliation{Kavli Nanoscience Institute and Thomas J. Watson, Sr., Laboratory of Applied Physics, California Institute of Technology, Pasadena, California 91125, USA}
\author{Mahmoud Kalaee} 
\affiliation{Kavli Nanoscience Institute and Thomas J. Watson, Sr., Laboratory of Applied Physics, California Institute of Technology, Pasadena, California 91125, USA}
\affiliation{Institute for Quantum Information and Matter, California Institute of Technology, Pasadena, California 91125, USA}
\author{Leyun Zang} 
\affiliation{Max Planck Institute for the Science of Light, G\"unther-Scharowsky-Stra\ss e 1/Bau 24, D-91058 Erlangen, Germany}
\author{Hannes Pfeifer} 
\affiliation{Max Planck Institute for the Science of Light, G\"unther-Scharowsky-Stra\ss e 1/Bau 24, D-91058 Erlangen, Germany}

\author{Florian Marquardt}
\affiliation{Max Planck Institute for the Science of Light, G\"unther-Scharowsky-Stra\ss e 1/Bau 24, D-91058 Erlangen, Germany}
\affiliation{Institute for Theoretical Physics, Department of Physics, Universit\"at Erlangen-N\"urnberg, 91058 Erlangen}

\author{Oskar Painter}
\affiliation{Kavli Nanoscience Institute and Thomas J. Watson, Sr., Laboratory of Applied Physics, California Institute of Technology, Pasadena, California 91125, USA}
\affiliation{Institute for Quantum Information and Matter, California Institute of Technology, Pasadena, California 91125, USA}
\email{opainter@caltech.edu}
\homepage{http://copilot.caltech.edu}

\begin{abstract}
We present the design, fabrication, and characterization of a planar silicon photonic crystal cavity in which large position-squared optomechanical coupling is realized.  The device consists of a double-slotted photonic crystal structure in which motion of a central beam mode couples to two high-$Q$ optical modes localized around each slot.  Electrostatic tuning of the structure is used to controllably hybridize the optical modes into supermodes which couple in a quadratic fashion to the motion of the beam.  From independent measurements of the anti-crossing of the optical modes and of the optical spring effect, the position-squared vacuum coupling rate is measured to be as large as $\gsqzpf/2\pi = 245$~Hz to the fundamental in-plane mechanical resonance of the structure at $\omegam/2\pi = 8.7$~MHz, which in displacement units corresponds to a coupling coefficient of $\gsq/2\pi = 1$~THz/nm$^2$.  This level of position-squared coupling is approximately five orders of magnitude larger than in conventional Fabry-Perot cavity systems.  
\end{abstract}

\maketitle

\section{Introduction}
\label{sec:intro}
In a cavity-optomechanical system the electromagnetic field of a resonant optical cavity or electrical circuit is coupled to the macroscopic motional degrees of freedom of a mechanical structure through radiation pressure~\cite{RevModPhys.86.1391}.  Cavity-optomechanical systems come in a multitude of different sizes and geometries, from cold atomic gases~\cite{PhysRevLett.105.133602} and nanoscale photonic structures~\cite{mattopto2009}, to the kilogram/kilometer scale interferometers developed for gravitational wave detection~\cite{LIGO-near-ground-state}.  Recent technological advancements in the field have led to the demonstration of optomechanically induced transparency~\cite{Safavi-Naeini2011-EIT,Weis2010-OMIT}, back-action cooling of a mechanical mode to its quantum ground state~\cite{OConnell2010-ground-state,Chan2011-ground-state-cooling,Teufel2011-cooling}, and ponderomotive squeezing of the light field~\cite{Brooks2012-squeezing,Safavi-Naeini2013-squeezing}. 

The interaction between light and mechanics in a cavity-optomechanical system is termed dispersive when it couples the frequency of the cavity to the position or amplitude of mechanical motion.  To lowest order this coupling is linear in mechanical displacement, however, the overall radiation pressure interaction is inherently nonlinear due to the dependence on optical intensity.  To date, this nonlinear interaction has been too weak to observe at the quantum level in all but the ultra-light cold atomic gases~\cite{PhysRevLett.105.133602}, and typically a large optical drive is used to parametrically enhance the optomechanical interaction.  Qualitatively novel quantum effects are expected when one takes a step beyond the standard linear coupling and exploits higher order dispersive optomechanical coupling.  In particular, ``$x^2$-coupling'' where the cavity frequency is coupled to the square of the mechanical displacement has been proposed as a means for realizing quantum non-demolition (QND) measurements of phonon number~\cite{Thompson_mim,Miao_standard_quantum_limit,Ludwig_enhanced}, measurement of phonon shot noise~\cite{Clerk_Phonon_Shot_Noise}, and the cooling and squeezing of mechanical motion~\cite{PhysRevA.77.033819,PhysRevA.82.021806,PhysRevA.89.023849}.  In addition to dispersive coupling, an effective $x^2$-coupling via optical homodyne measurement has also been proposed, with the capability of generating and detecting non-Gaussian motional states~\cite{Vanner_prX_2011}.

The dispersive $x^2$-coupling between optical and mechanical resonator modes in a cavity-optomechanical system is described by the coefficient $\gsq \equiv \partial^2 \omegac/\partial x^2$, where $\omegac$ is the frequency of the optical resonance of interest and $x$ is the generalized amplitude coordinate of the displacement field of the mechanical resonance.  One can show via second-order perturbation theory~\cite{Amir_thesis2013,Kaviani2014-paddle} that $x^2$-coupling arises due to linear cross-coupling between the optical mode of interest and other modes of the cavity.  In the case of two nearby resonant modes, the magnitude of the $x^2$-coupling coefficient depends upon the square of the magnitude of the linear cross-coupling between the two modes ($\glin$) and inversely on their frequency separation or tunnel coupling rate ($2J$), $\gsq = \glin^2/2J$.  In pioneering work by Thompson, et al.~\cite{Thompson_mim}, a Fabry-Perot cavity with an optically-thin \SiN membrane positioned in between the two end mirrors was used to realize $x^2$-coupling via hybridization of the degenerate modes of optical cavities formed on either side of the partially reflecting membrane.  More recently, a number of cavity-optomechanical systems displaying $x^2$-coupling have been explored, including double microdisk resonators~\cite{Hill2013}, microdisk-cantilever systems~\cite{PhysRevA.89.053838}, microsphere-nanostring systems~\cite{Brawley2014-nanostring}, atomic gases trapped in Fabry-Perot cavities~\cite{PhysRevLett.105.133602}, and paddle nano-cavities~\cite{Kaviani2014-paddle}. 
 
Despite significant technical advances made in recent years~\cite{Flowers-jacobs_fiber,Sankey_tuning,PhysRevA.89.053838,Kaviani2014-paddle}, the use of $x^2$-coupling for measuring or preparing non-classical quantum states of a mesoscopic mechanical resonator remains an elusive goal.  This is a direct result of the small coupling rate to motion at the quantum level, which for $x^2$-coupling scales as the square of the zero-point motion amplitude of the mechanical resonator, $\xzpf^2 = \hbar/2\meff\omegam$, where $\meff$ is the motional mass of the resonator and $\omegam$ is the resonant frequency.  As described in Ref.~\cite{Ludwig_enhanced}, one method to greatly enhance the $x^2$-coupling in a multi-mode cavity-optomechanical system is to fine tune the mode splitting $2J$ to that of the mechanical resonance frequency.  

In this work we utilize a quasi two-dimensional (2D) photonic crystal structure to create an optical cavity supporting a pair of high-$Q$ optical resonances in the $1500$~nm wavelength band exhibiting large linear optomechanical coupling.  The double-slotted structure is split into two outer slabs and a central nanobeam, all three of which are free to move, and electrostatic actuators are integrated into the outer slabs to allow for both the trimming of the optical modes into resonance and tuning of the tunnel coupling rate $J$.  Due to the form of the underlying photonic bandstructure the spectral ordering of the cavity supermodes in this structure may be reversed, enabling arbitrarily small values of $J$ to be realized.  Measurement of the optical resonance anti-crossing curve, along with calibration of the linear optomechanical coupling through measurement of the dynamic optical spring effect, yields an estimated $x^2$-coupling coefficient as large as $\gsq/2\pi = 1$~THz/nm$^2$ ($\gsqzpf/2\pi = 245$~Hz) to the fundamental mechanical resonance of the central beam at $\omegam/2\pi=8.7$~MHz.  Additional measurements of $\gsq$ through the dynamic and static optical spring effects are also presented.  This level of $x^2$-coupling is approximately five orders of magnitude larger than in conventional Fabry-Perot MIM systems demonstrated to date~\cite{Sankey_tuning}, and two orders of magnitude larger than in the smaller mode volume fiber-gap cavities~\cite{Flowers-jacobs_fiber}. 



\section{Theoretical Background}
\label{sec:theory}

\begin{figure}[!t]
 \begin{center}
 \includegraphics[width=\columnwidth]{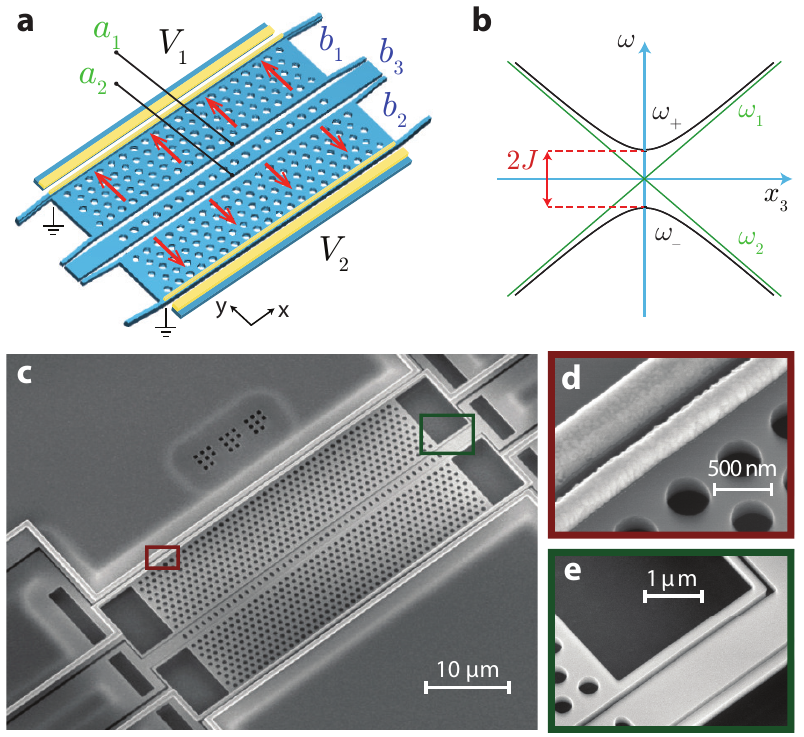}
 \caption{(a) Double-slotted photonic crystal cavity with optical cavity resonances ($a_1,a_2$) centered around the two slots, and three fundamental in-plane mechanical resonances corresponding to motion of the outer slabs ($b_1,b_2$) and the central nanobeam ($b_3$). Tuning the equilibrium position of the outer slabs $b_1$ and $b_2$, and consequently the slot size on either side of the central nanobeam, is achieved by pulling on the slabs (red arrows) through an electrostatic force proportional to the square of the voltage applied to capacitors on the outer edge of each slab. (b) Dispersion of the optical modes as a function of $x_3$, the in-plane displacement of the central nanobeam from its symmetric equilibrium position.  Due to tunnel coupling at a rate $J$ the slot modes $a_1$ and $a_2$ hybridize into the even and odd supermodes $a_{+}$ and $a_{-}$, which have a parabolic dispersion near the central anti-crossing point ($\omega_{1}=\omega_{2}$).  (c) SEM image of a fabricated double-slotted photonic crystal device in the SOI material system. (d) Zoom-in SEM image showing the capacitor gap ($\sim 100$~nm) for the capacitor of one of the outer slabs. (e) Zoom-in SEM image showing some of the suspending tethers of the outer slabs which are of length $2.5$~$\mu$m and width $155$~nm.  The central beam, which is much wider, is also shown in this image.\label{fig_sample}}
\end{center}
\end{figure}

Before we discuss the specific double-slotted photonic crystal cavity-optomechanical system studied in this work, we consider a more generic multi-moded system consisting of two optical modes which are dispersively coupled to the same mechanical mode, and in which the dispersion of each mode is linear with the amplitude coordinate $x$ of the mechanical mode.  If we further assume a purely optical coupling between the two optical modes, the Hamiltonian for such a three-mode optomechanical system in the absence of drive and dissipation is given by $\hat{H}= \Hnot + \HOM + \Hint$:

\begin{align}
&\Hnot=\hbar \omega_{1} \hat{a}_1^\dagger \hat{a}_1 + \hbar \omega_{2} \hat{a}_2^\dagger \hat{a}_2 + \hbar \omegamnot \hat{b}^\dagger \hat{b}, \\
&\HOM= \hbar (\glin_{1}\hat{a}_1^\dagger \hat{a}_1+\glin_{2}\hat{a}_2^\dagger \hat{a}_2)\hat{x}\label{eq:om:slots},\\
&\Hint=\hbar J (\hat{a}_1^\dagger \hat{a}_2+\hat{a}_2^\dagger \hat{a}_1).
\end{align} 

\noindent Here, $\hat{a}_{i}$ and $\omega_{i}$ are the annihilation operator and the bare resonance frequency of the $i$th optical resonance, $\hat{x}=(\hat{b}^\dagger+\hat{b})\xzpf$ is the quantized amplitude of motion, $\xzpf$ the zero point amplitude of the mechanical resonance, $\omegamnot$ is the bare mechanical resonance frequency, and $\glin_i$ is the linear optomechanical coupling constant of the $i$th optical mode to the mechanical resonance.  Without loss of generality, we take the bare optical resonance frequencies to be equal ($\omega_1=\omega_2 \equiv \omega_0$), allowing us to rewrite the Hamiltonian in the normal mode basis $\hat{a}_{\pm}=(\hat{a}_1\pm \hat{a}_2)/\sqrt{2}$ as, 

\begin{multline}\label{eq:H_pm_basis}
\hat{H}=\hbar\omega_{+}(0)\hat{a}_+^\dagger \hat{a}_+ + \hbar\omega_{-}(0)\hat{a}_-^\dagger \hat{a}_- + \hbar \omegamnot \hat{b}^\dagger \hat{b} \\
+ \hbar \left(\frac{g_1+g_2}{2}\right)\left(\hat{a}_+^\dagger \hat{a}_+ + \hat{a}_-^\dagger \hat{a}_-\right)\hat{x} \\
+ \hbar \left(\frac{g_1-g_2}{2}\right)\left(\hat{a}_+^\dagger \hat{a}_- + \hat{a}_-^\dagger \hat{a}_+\right)\hat{x},
\end{multline} 

\noindent where $\omega_{\pm}(0) = \omega_0 \pm J$.

For $|J| \gg \omega_m$ such that $\hat{x}$ can be treated as a quasi-static variable~\cite{Ludwig_enhanced,Miao_standard_quantum_limit}, the Hamiltonian can be diagonalized resulting in eigenfrequencies $\omega_\pm(\hat{x})$,

\begin{equation}\label{eq:omega_pm}
\omega_\pm(\hat{x}) \approx \omega_0+\frac{(g_1+g_2)}{2}\hat{x} \pm J\left( 1+ \frac{(g_1-g_2)^2}{8J^2} \hat{x}^2 \right).
\end{equation}

\noindent As shown below, in the case of the fundamental in-plane motion of the outer slabs of the double-slotted photonic crystal cavity we have only one of $g_{1}$ or $g_{2}$ non-zero, whereas in the case of the fundamental in-plane motion of the central nanobeam we have $g_{1} \approx -g_{2}$.

For a system in which the mechanical mode couples to the $a_{1}$ and $a_{2}$ optical modes with linear dispersive coupling of equal magnitude but opposite sign ($g_1=-g_2=g$), the dispersion in the quasi-static normal mode basis is purely quadratic with effective $x^2$-coupling coefficient,

\begin{equation}\label{eq:g'}
\gsq=\glin^2/2J,
\end{equation}

\noindent and quasi-static Hamiltonian, 

\begin{multline}\label{eq:H_AS_QS}
\hat{H} \approx \hbar\left(\omega_{+}(0)+\gsq \hat{x}^2\right)\hat{n}_{+} \\ 
+ \hbar\left(\omega_{-}(0)-\gsq \hat{x}^2\right)\hat{n}_{-} + \hbar \omegamnot \hat{n}_{b},
\end{multline}
        
\noindent where $\hat{n}_{\pm}$ are the number operators for the $a_{\pm}$ supermodes and $\hat{n}_{b}$ is the number operator for the mechanical mode.  Rearranging this equation slightly highlights the interpretation of the $x^2$ optomechanical coupling as inducing a static optical spring,

\begin{multline}\label{eq:H_AS_QS_spring}
\hat{H} \approx \hbar\omega_{+}(0)\hat{n}_{+} + \hbar\omega_{-}(0)\hat{n}_{-} \\
+ \hbar \left[\omegamnot \hat{n}_{b} + \gsq \left(\hat{n}_{+} - \hat{n}_{-}\right)\hat{x}^2\right],
\end{multline}

\noindent where the static optical spring constant $\bar{k}_{s} = 2\hbar\gsq\left(n_{+} - n_{-}\right)$ depends upon the average intra-cavity photon number in the even and odd optical supermodes, $n_{\pm} \equiv \langle \hat{n}_{\pm} \rangle$.
 
For a sideband resolved system ($\omega_m \gg \kappa$), the quasi-static Hamiltonian can be further approximated using a rotating-wave approximation as,

\begin{multline}\label{eq:HRWA}
\hat{H} \approx \hbar[\omega_{+}(0) + 2\gsqzpf(\hat{n}_{b} + 1/2)]\hat{n}_{+} \\
+ \hbar[\omega_{-}(0) - 2\gsqzpf(\hat{n}_{b} + 1/2)]\hat{n}_{-} + \hbar \omegamnot \hat{n}_{b},
\end{multline}

\noindent where $\gsqzpf \equiv \gsq\xzpf^2 = \glinzpf^2/2J$ and $\glinzpf \equiv \glin\xzpf$ are the $x^2$ and linear vacuum coupling rates, respectively. It is tempting to assume from eq.~(\ref{eq:HRWA}) that by monitoring the optical transmission through the even or odd supermode resonances, that one can then perform a continuous quantum non-demolition (QND) measurement of the phonon number in the mechanical resonator~\cite{Santamore2004,Thompson_mim,Jayich2008,Gangat_Milburn}.  As noted in Refs.~\cite{Ludwig_enhanced,Miao_standard_quantum_limit}, however, the quasi-static picture described by the dispersion of eq.~(\ref{eq:omega_pm}) fails to capture residual effects resulting from the non-resonant scattering between the $a_{+}$ and $a_{-}$ supermodes which depends linearly on $\hat{x}$ (last term of eq.~(\ref{eq:H_pm_basis})).  Only in the vacuum strong coupling limit ($\glinzpf/\kappa \gtrsim 1$) can one realize a QND measurement of phonon number~\cite{Ludwig_enhanced,Miao_standard_quantum_limit}.

The regime of $|2J| \sim \omegamnot$ is also very interesting, and explored in depth in Refs.~\cite{Ludwig_enhanced,Stannigel_Lukin}.  Transforming to a reference frame which removes in eq.~(\ref{eq:H_pm_basis}) the radiation pressure interaction between the even and odd supermodes to first order in $\glin$, yields an effective Hamiltonian given by~\cite{Ludwig_enhanced,Ludwig_thesis2013},

\begin{multline}\label{eq:Heff}
\hat{H}_{\text{eff}} \approx \hbar\omega_{+}(0)\hat{n}_{+} + \hbar\omega_{-}(0)\hat{n}_{-} + \hbar\omegamnot \hat{n}_{b}  \\
+ \hbar\frac{\glinzpf^2}{2}\left[\frac{1}{2J-\omegamnot} + \frac{1}{2J+\omegamnot}\right]\left(\hat{a}^{\dagger}_{+}\hat{a}_{+} - \hat{a}^{\dagger}_{-}\hat{a}_{-} \right)\left(\hat{b} + \hat{b}^{\dagger}\right)^2 \\
+ \hbar\frac{\glinzpf^2}{2}\left[\frac{1}{2J-\omegamnot} - \frac{1}{2J+\omegamnot}\right]\left(\hat{a}^{\dagger}_{+}\hat{a}_{-} + \hat{a}^{\dagger}_{-}\hat{a}_{+} \right)^2,
\end{multline} 

\noindent where we assume $|\glinzpf/\delta| \ll 1$ for $\delta \equiv |2J| - \omegamnot$, and terms of order $\glinzpf^3/(2J \pm \omegamnot)^2$ and higher have been neglected.  In the limit $|J| \gg \omegamnot$ we recover the quasi-static result of eq.~(\ref{eq:H_AS_QS}), whereas in the near-resonant limit of $|\delta| \ll |J|,\omegamnot$ we arrive at,

\begin{multline}\label{eq:Heffres}
\hat{H}_{\text{eff}} \approx \hbar\omega_{+}(0)\hat{n}_{+} + \hbar\omega_{-}(0)\hat{n}_{-} + \hbar\omegamnot \hat{n}_{b}  \\
+ \hbar\frac{\glinzpf^2}{2\delta}\big[2\sgn(J)\left(\hat{n}_{+}-\hat{n}_{-}\right)\left(\hat{n}_{b}+1\right) \\
+ 2\hat{n}_{+}\hat{n}_{-} + \hat{n}_{+} + \hat{n}_{-}\big].
\end{multline}

\noindent Here we have neglected highly oscillatory terms such as $(\hat{a}_{+}^{\dagger}\hat{a}_{-})^2$ and $\hat{b}^2$, a good approximation in the sideband resolved regime ($\kappa \ll \omegamnot, |J|$).  From eq.~(\ref{eq:Heffres}) we find that the frequency shift per phonon of the optical resonances is much larger than in the quasi-static case ($\glinzpf^2/2|\delta| \gg \glinzpf^2/2|J|$).  Although a QND measurement of phonon number still requires the vacuum strong coupling limit, this enhanced read-out sensitivity is attainable even for $\glinzpf/\kappa \ll 1$.   Equation~(\ref{eq:Heffres}) also indicates that, much like the QND measurement of phonon number, in the near-resonant limit a measurement of the intra-cavity photon number stored in one optical supermode can be performed by monitoring the transmission of light through the other supermode~\cite{Ludwig_enhanced,Ludwig_thesis2013}.

\section{Double-Slotted Photonic Crystal Optomechanical Cavity}
\label{sec:PC_cavity}
A sketch of the double-slotted photonic crystal cavity structure is shown in Fig.~\ref{fig_sample}a.  As detailed below and elsewhere~\cite{Kalaee2015}, the optical cavity structure can be thought of as formed from two coupled photonic crystal waveguides, one around each of the nanoscale slots, and each with propagation direction along the $z$-axis.  A small adjustment ($\sim 5\%$) in the lattice constant is used to produce a local shift in the waveguide band-edge frequency, resulting in trapping of optical resonance to this ``defect'' region.  Optical tunneling across the central photonic crystal beam, which in this case contains only a single row of holes, couples the cavity mode of slot 1 ($a_1$) to the cavity mode of slot 2 ($a_2$).

The two outer photonic crystal slabs and the central nanobeam are all mechanically compliant, behaving as independent mechanical resonators.  The mechanical resonances of interest in this work are the fundamental in-plane flexural modes of the top slab, the bottom slab, and the central nanobeam, denoted by $b_1$, $b_2$ and $b_3$, respectively. For a perfectly symmetric structure about the $z$-axis of the central nanobeam, the linear dispersive coupling coefficients of the $b_3$ mode of the central nanobeam to the two slot modes $a_1$ and $a_2$ are equal in magnitude but opposite in sign, resulting in a vanishing linear coupling at the resonant point where $\omega_{1}=\omega_{2}$ (c.f., eq.~(\ref{eq:omega_pm})). Figure~\ref{fig_sample}b shows a plot of the dispersion of the optical resonances as a function of the nanobeam's in-plane displacement ($x_3$), illustrating how the linear dispersion of the slot modes ($a_1,a_2$) transforms into quadratic dispersion for the upper and lower supermode branches ($a_{+},a_{-}$) in the presence of tunnel coupling $J$. The mechanical modes of the outer slabs ($b_1,b_2$) provide degrees of freedom for post-fabrication tuning of the slotted waveguide optical modes, i.e., to symmetrize the structure such that $\omega_{1}=\omega_{2}$.  This is achieved in practice by integrating metallic electrodes which form capacitors at the outer edge of the two slabs of the structure as schematically shown in Fig.~\ref{fig_sample}a.

The double-slotted photonic crystal cavity of this work is realized in the silicon-on-insulator (SOI) material system, with a top silicon device layer thickness of $220$~nm and an underlying buried oxide (BOX) layer of $3$~$\mu$m.  Fabrication begins with the patterning of the metal electrodes of the capacitors, and involves electron beam (ebeam) lithography followed by evaporation and lift-off of a bi-layer consisting of a $5$~nm sticking layer of chromium and a $150$~nm layer of gold.  After lift-off we deposit uniformly a $\sim 4$~nm protective layer of silicon dioxide.  A second electron beam lithography step is performed, aligned to the first, to form the pattern of the photonic crystal and the nanoscale slots which separate the central nanobeam from the outer slabs.  At this step, we also pattern the support tethers of the outer slabs and the cut lines which define and isolate the outer capacitors.  A fluorine based (C$_4$F$_8$ and SF$_{6}$) inductively coupled reactive-ion etch (ICP-RIE) is used to transfer the ebeam lithography pattern through the silicon device layer.  The remaining ebeam resist is stripped using trichloroethylene, and then the sample is cleaned in a heated Piranha (H$_2$SO$_4$:H$_2$O$_2$) solution.  The devices are then released using a hydrofluoric (HF)acid etch to remove the sacrificial BOX layer (this also removes the deposited protective silicon dioxide layer), followed by a water rinse and critical point drying.  

A scanning electron microscope (SEM) image showing the overall fabricated device structure is shown in Fig.~\ref{fig_sample}c.  Zoom-ins of the capacitor region of one of the outer slabs and the tether region at the end of the nanobeam are shown in Figs.~\ref{fig_sample}d and e, respectively.  Note that the geometry of the capacitors and the stiffness of the support tethers determine how tunable the structure is under application of voltages to the capacitor electrodes.  The outermost electrode of each slab is connected to an independent low-noise DC voltage source, while the innermost electrodes are connected to a common ground, thereby allowing one to independently pull on each outer slab with voltages $V_{1}$ and $V_{2}$.  In this configuration, we are limited to increasing the slots defining the optical modes around the central nanobeam.   

\subsection{Photonic bandstructure}
\label{subsec:photonic_bandstructure}
 
\begin{figure}[!t]
\begin{center}
\includegraphics[width=\columnwidth]{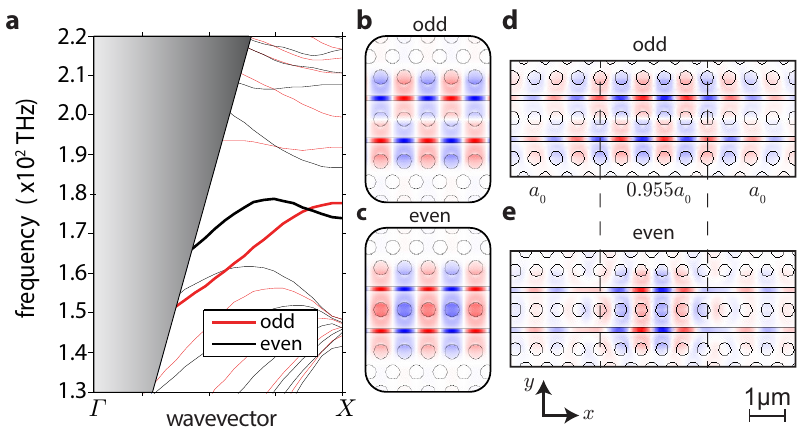}
\caption{(a) Bandstructure diagram of the periodic (along $x$) double-slotted photonic crystal waveguide structure.  Here we only show photonic bands that are composed of modes with even vector symmetry around the ``vertical'' ($\sigma_{z}$) mirror plane.  The two waveguide bands of interest lie within the quasi-2D photonic bandgap of the outer photonic crystal slabs and are shown as bold red and black curves.  These waveguide bands are labeled ``even'' (bold black curve) and ``odd'' (bold red curve) due to the spatial symmetry of their mode shape for the dominant electric field polarization in the $y$-direction, $E_{y}$.  The simulated structure is defined by the lattice constant between nearest neighbor holes in the hexagonal lattice ($a_{0}=480$~nm), the thickness of the silicon slab ($d=220$~nm), the width of the two slots ($s=100$~nm), and the refractive index of the silicon layer ($n_{\text{Si}} = 3.42$).  The hole radius in the outer slabs and the central nanobeam is $r=144$~nm.  The grey shaded region represents a continuum of radiation modes which lie above the light cone for the air cladding which surrounds the undercut silicon slab structure. (b) Normalized $E_y$ field profile at the $X$-point of the odd waveguide supermode, shown for several unit cells along the $x$ guiding axis. (c) $E_y$ field profile of the even waveguide supermode.  Waveguide simulations of (a-c) were performed using the plane-wave mode solver MPB~\cite{Johnson2001,MPB}.  Normalized $E_y$ field profile of the corresponding localized cavity supermodes of (d) odd and (e) even spatial symmetry about the horizontal mirror plane. The lattice constant $a_0$ is decreased by $4.5\%$ for the central five lattice constants between the dashed lines to localize the waveguide modes.  Simulations of the full cavity modes were performed using the COMSOL finite-element method mode solver package~\cite{COMSOL}.\label{fig_bandstructure}}
\end{center}
\end{figure}

To further understand the optical properties of the double-slotted photonic crystal cavity, we display in Fig.~\ref{fig_bandstructure}a the photonic bandstructure of the periodic waveguide structure. The parameters of the waveguide are given in the caption of Fig.~\ref{fig_bandstructure}a.  Here we only show photonic bands that are composed of waveguide modes with even vector symmetry around the ``vertical'' mirror plane ($\sigma_{z}$), where the vertical mirror plane is defined by the $z$-axis normal and lies in the middle of the thin-film silicon slab.   The fundamental (lowest lying) optical waveguide bands are of predominantly transverse (in-plane) electric field polarization, and are thus called TE-like.  In the case of a perfectly symmetric structure, we can further classify the waveguide bands by their odd or even symmetry about the ``horizontal'' mirror plane ($\sigma_{y}$) defined by the $y$-axis normal and cutting through the middle of the central nanobeam.  The two waveguide bands of interest that lie within the quasi-2D photonic bandgap of the outer photonic crystal slabs, shown as bold red and black curves, are labeled ``even'' and ``odd'' depending on the spatial symmetry with respect to $\sigma_{y}$ of their mode shape for the dominant electric field polarization in the $y$-direction, $E_{y}$ (note that this labeling is opposite to their vector symmetry).  The $E_{y}$ spatial mode profiles at the $X$-point for the odd and even waveguide supermodes are shown in Figs.~\ref{fig_bandstructure}b and c, respectively.  

An optical cavity is defined by decreasing the lattice constant $4.5\%$ below the nominal value of $a_{0}=480$~nm for the middle five periods of the waveguide (see Fig.~\ref{fig_bandstructure}d).  This has the effect of locally pushing the bands toward higher frequencies~\cite{Safavi-Naeini_APL_2010,Winger:chip-scale}, which creates an effective potential that localizes the optical waveguide modes along the $x$-axis of the waveguide.  The resulting odd and even TE-like cavity supermodes are shown in Figs.~\ref{fig_bandstructure}d and e, respectively.  These optical modes correspond to the normal modes $a_{+}$ and $a_{-}$ in Section~\ref{sec:theory}, which are symmetric and anti-symmetric superpositions, respectively, of the cavity modes localized around each slot ($a_{1}$ and $a_{2}$).  Due to the non-monotonic decrease in the even waveguide supermode as one moves away from the $X$-bandedge (c.f., Fig.~\ref{fig_bandstructure}a), we find that the simulated optical $Q$-factor of the even $a_{+}$ cavity supermode is significantly lower than that of the odd $a_{-}$ cavity supermode.  This will be a key distinguishing feature found in the measured devices as well. 

\subsection{Optical tuning simulations}
\label{subsec:numerical_tuning}

\begin{figure}[!t]
\begin{center}
\includegraphics[width=\columnwidth]{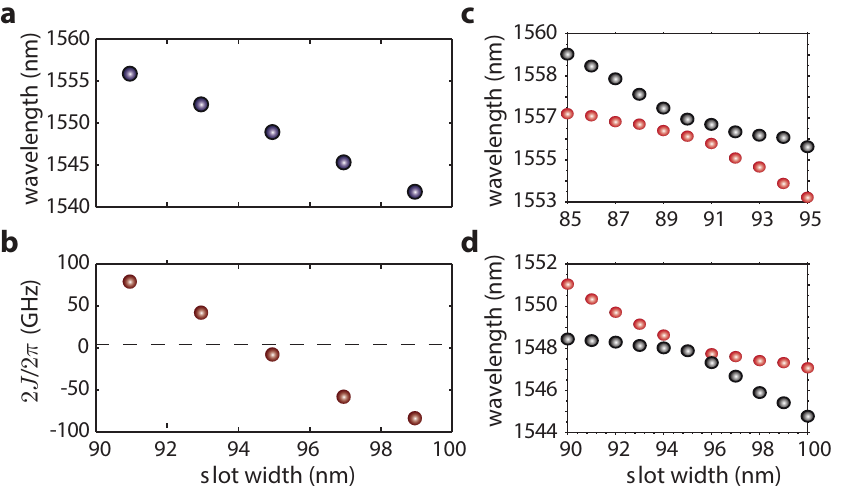}
\caption{Symmetric tuning of the slot widths of the double-slotted photonic crystal cavity showing (a) the mean wavelength shift and (b) the splitting $2J = \omega_+-\omega_-$ of the even and odd cavity supermodes versus slot width $s=s_{1}=s_{2}$.  (c-d)  Avoided crossing of the cavity supermodes obtained by tuning $s_1$ while keeping $s_2$ fixed at (c) $s_2=90$~nm and (d) $s_2=95$~nm.  For all simulations in (a-d) the parameters of the cavity structure are the same as in Fig.~\ref{fig_bandstructure}, except for the slot widths.  The simulations were performed using the COMSOL FEM mode solver~\cite{COMSOL}.\label{fig_symmetry_breaking}}
\end{center}
\end{figure}

The slot width in the simulated waveguide and cavity structures of Fig.~\ref{fig_bandstructure} is set at $s=100$~nm.  For this slot width we find a lower frequency for the even ($a_{+}$) supermode than the odd ($a_{-}$) supermode at the $X$-point photonic bandedge of the periodic waveguide and in the case of the localized cavity modes.  Figure~\ref{fig_symmetry_breaking} presents finite-element method (FEM) simulations of the optical cavity for slot sizes swept from $90$~nm to $100$~nm in steps of $1$~nm, all other parameters the same as in Fig.~\ref{fig_bandstructure}.  For the slot widths tuned symmetrically ($s_1=s_2=s$), the mean wavelength of the even and odd cavity supermodes and their frequency splitting $2J=\omega_+-\omega_-$ are plotted in Fig.~\ref{fig_symmetry_breaking}a and Fig.~\ref{fig_symmetry_breaking}b, respectively. As expected the mean wavelength drops for increasing slot width.  The frequency splitting, however, also monotonically decreases with slot width, going from a positive value for $s=90$~nm to a negative for $s=100$~nm slots and crossing zero for a slot width of $s=95$~nm.  In Figs.~\ref{fig_symmetry_breaking}c and d the symmetry is broken by keeping $s_2$ fixed and scanning $s_1$; the cavity supermodes are driven through an anti-crossing with a splitting determined by the fixed slot width $s_{2}$.

The spectral inversion of the even $a_+$ and odd $a_-$ cavity supermodes predicted in Fig.~\ref{fig_symmetry_breaking}b originates in the unequal overlap of each mode with the air slots separating the two outer slabs from the central nanobeam.  The odd supermode tends to be pushed further from the middle of the central nanobeam, having slightly larger overlap with the air slots.  An increase in the air region for increased slot size leads to a blue shift of both cavity supermodes. The odd mode having a larger electric field energy density in the air slots than the even mode is more affected by a change in the slot widths. Therefore, upon equal increase of the slot widths, the odd mode experiences larger frequency shifts than the even mode, which results in a tuning of the frequency splitting. For particular geometrical parameters of the central nanobeam~\cite{Kalaee2015}, a change in the slot widths is sufficient to invert the spectral ordering of the supermodes. This means that arbitrarily small splittings can potentially be realized, which is important for applications in $x^2$ detection where the splitting enters inversely in the coupling (for the quasi-static case).
 
\section{Experimental measurements}
\label{sec:exp_results}

\begin{figure*}[!t]
\begin{center}
\includegraphics[width=\textwidth]{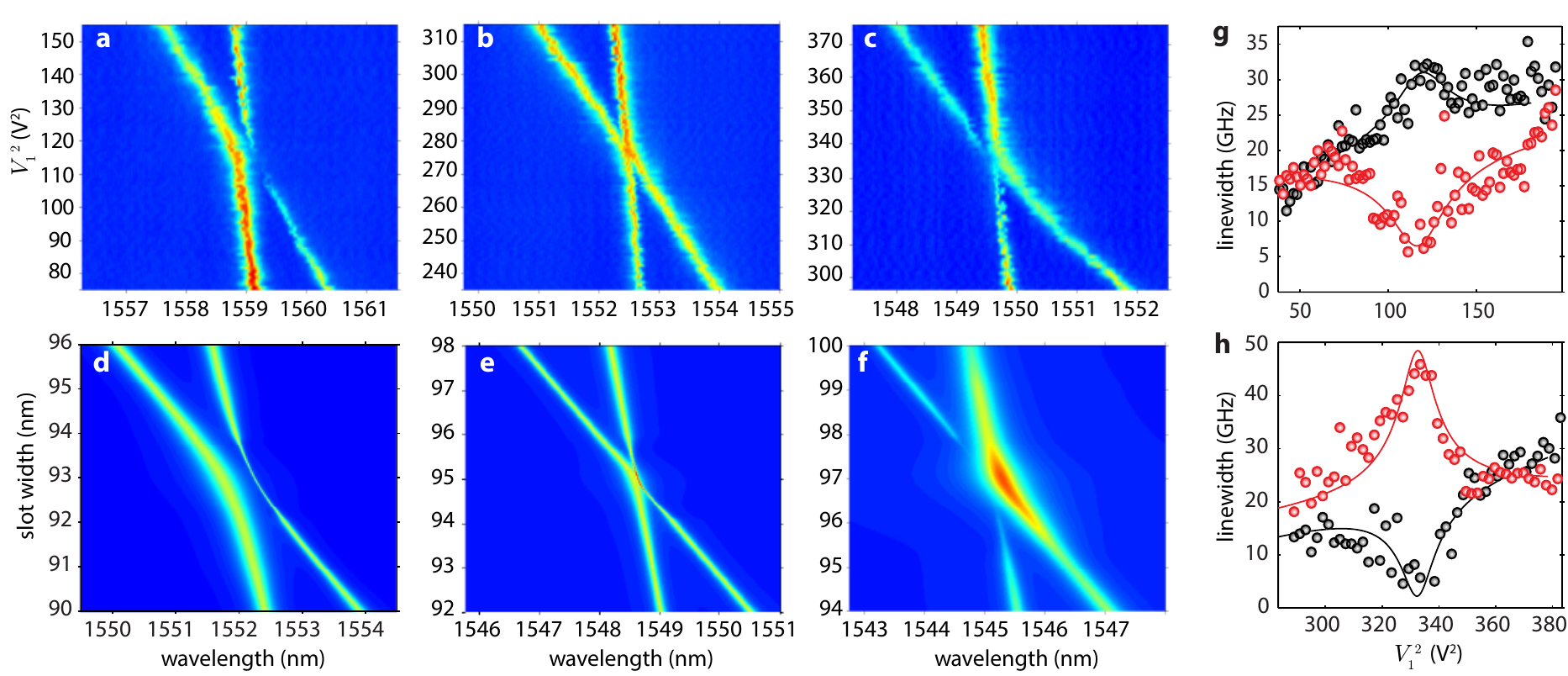}
\caption{(a-c) Optical transmission measurements versus the wavelength of the probe laser showing the cavity mode anti-crossing and tuning of the photon tunneling rate. In these measurements the probe laser wavelength (horizontal axis) is scanned across the optical cavity resonances as the voltage across the first capacitor $V_{1}$ is swept from low to high (vertical axis shows $V^2_{1}$ in V$^2$, proportional to slab displacement).  The second capacitor is held fixed at (a) $V_{2}=1$V, (b) $V_{2}=15$V and (c) $V_{2}=18$V.  The colorscale indicates the fractional change in the optical transmission level, $\Delta T$, with blue corresponding to $\Delta T = 0$ and red corresponding to $\Delta T \approx 0.25$. From the three anti-crossing curves we measure a splitting $2J$ equal to (a) $50$~GHz, (b) $12$~GHz, and (c) $-25$~GHz.  (d-f)  Corresponding simulations of the normalized optical transmission spectra for the slot width $s_1$ varied and the second slot width held fixed at (d) $s_2=93$~nm, (e) $s_2=95$~nm and (f) $s_2=97$~nm. The dispersion and tunneling rate of the slot modes are taken from simulations similar to that found in Fig.~\ref{fig_symmetry_breaking}.  (g)  and (h) show the measured linewidths of the high frequency upper (black) and low frequency lower (red) optical resonance branches as a function of $V^2_{1}$, extracted from (a) and (c), respectively. The narrowing (broadening) is a characteristic of the odd (even) nature of the cavity supermode, indicating the inversion of the even and odd supermodes for the two voltage conditions $V_{2}=1$~V and $V_{2}=18$~V. The lines are guides for the eye. \label{fig_tuning_small_slot}}
\end{center}
\end{figure*}

Optical testing of the fabricated devices is performed in a nitrogen-purged enclosure at room temperature and pressure.  A dimpled optical fiber taper is used to locally excite and collect light from the photonic crystal cavity, details of which can be found in Ref.~\cite{Michael:07}. The light from a tunable, narrow-bandwidth laser source in the telecom $1550$~nm wavelength band (New Focus, Velocity series) is evanescently coupled from the fiber taper into the device with the fiber taper guiding axis parallel with that of the photonic crystal waveguide axis, and the fiber taper positioned laterally at the center of the nanobeam and vertically a few hundreds of nanometers above the surface of the silicon chip.  Relative positioning of the fiber taper to the chip is accomplished using a multi-axis set of encoded DC-motor stages with $50$~nm step resolution.  The polarization of the light in the fiber is polarized parallel with the surface chip in order to optimize the coupling to the in-plane polarization of the cavity modes.  

With the taper placed suitably close to a photonic crystal cavity ($\sim 200$~nm), the transmission spectrum of the laser probe through the device features resonance dips at the supermode resonance frequencies, as shown in the intensity plots of Figs.~\ref{fig_tuning_small_slot}a-c.   The resonance frequencies of the cavity modes are tuned via displacement of the top and bottom photonic crystal slabs, which can be actuated independently using their respective capacitor voltages $V_{1}$ and $V_{2}$. The capacitive force is proportional to the applied voltage squared~\cite{Winger:chip-scale}, and thus increasing the voltage $V_{i}$ on a given capacitor widens the waveguide slot $s_{i}$ and (predominantly) increases the slot mode frequency $a_i$ (note the other optical slot mode frequency also increases slightly). For the devices studied in this work, the slab tuning coefficient with applied voltage ($\alpha_\textnormal{cap}$) is estimated from SEM analysis of the resulting structure dimensions and FEM electromechanical simulations to be $\alpha_\textnormal{cap}=25$~pm/V$^2$. 

We fabricated devices with slot widths targeted for a range of $75$-$85$~nm, chosen smaller than the expected zero-splitting slot width of $s=95$~nm so that the capacitors could be used to tune through the zero-splitting point. While splittings larger than $150$~GHz were observed in the nominal $85$~nm slot width devices, splittings as small as $10$~GHz could be resolved in the smaller $75$~nm slot devices.  As such, in what follows we focus on the results from a single device with as-fabricated slot size of $s \approx 75$~nm.


\subsection{Anti-crossing measurements}
\label{subsec:anticrossing_measurements}

Figure~\ref{fig_tuning_small_slot} shows intensity plots of the normalized optical transmission through the optical fiber taper when evanescently coupled to the photonic crystal cavity of a device with nominal slot width $s=75$~nm.  Here a series of optical transmission spectrum are measured by sweeping the probe laser frequency and the voltage $V_{1}$, with $V_{2}$ fixed at three different values.  The estimated anti-crossing splitting from the measured dispersion of the cavity supermodes is $2J/2\pi = 50$~GHz, $12$~GHz, and $-25$~GHz for $V_{2} = 1$~V, $15$~V, and $18$~V, respectively.  In order to distinguish between the odd and even cavity supermodes at the anti-crossing point, we use the fact that both the coupling rate to the fiber taper $\kappa_e$ and the intrinsic linewidth $\kappa_i$ depend upon the symmetry of the cavity mode. First, the odd supermode branch becomes dark at the anti-crossing because it cannot couple to the symmetric fiber taper mode. Second, in the vicinity of the anti-crossing point the linewidth of the odd supermode branch narrows while the linewidth of the even supermode branch broadens~\cite{Kalaee2015}.  Far from the anti-crossing region, the branches are asymptotic to individual slot modes and their linewidths and couplings to the fiber taper are similar.  

These features are clearly evident in the optical transmission spectra of Figs.~\ref{fig_tuning_small_slot}a-c, as well as in the measured linewidth of the optical supermode resonances shown in Figs.~\ref{fig_tuning_small_slot}g-h. Figure~\ref{fig_tuning_small_slot}a was taken with a small voltage $V_{2}=1$~V, corresponding to a small slot width at the anti-crossing point, and thus consistent with the even mode frequency being higher than the odd mode frequency for small slot widths (c.f., Fig.~\ref{fig_symmetry_breaking}b).  The exact opposite identification is made in Fig.~\ref{fig_tuning_small_slot}c where $V_{2}=18$~V is much larger, corresponding to a larger slot width at the anti-crossing point. Fig.~\ref{fig_tuning_small_slot}b with $V_{2}=15$~V  is close to the zero-splitting condition.  For comparison, a simulation of the expected anti-crossing curves are shown in Figs.~\ref{fig_tuning_small_slot}d, e, and f for $s_{2}=93$, $95$, and $97$~nm, respectively. Here we have taken the even superposition of the slot modes to have a lower $Q$-factor than the odd superposition of the slot modes, and the coupling of the fiber taper to be much stronger to the even mode than the odd mode, consistent with results from numerical FEM simulations.  Good qualitative correspondence is found with the measured transmission curves of Figs.~\ref{fig_tuning_small_slot}a-c.             
 
An estimate of the $x^2$-coupling coefficient $\gsq_{b_3}$ can found from the simulated value of $\alpha_\textnormal{cap}$ and a fit to the measured tuning curves of Fig.~\ref{fig_tuning_small_slot} away from the anti-crossing point.  Consider the anti-crossing curve of Fig.~\ref{fig_tuning_small_slot}b with the smallest splitting.  Far from the anti-crossing point the tuning of the $a_{1}$ and $a_{2}$ slot modes are measured to be linear with the square of $V_{1}$: $\glin_{a_1,V^2_1}/2\pi = 3.9$~GHz/V$^2$ and $\glin_{a_2,V^2_1}/2\pi = 0.5$~GHz/V$^2$.  For the simulated value of $\alpha_\textnormal{cap} = 0.025$~nm/V$^2$ the corresponding linear dispersive coefficients versus the first slot width are $\glin_{a_1,\delta s_1}/2\pi = 156$~GHz/nm and $\glin_{a_2,\delta s_1}/2\pi = 20$~GHz/nm.  Noting that a displacement amplitude $x_{3}$ for the fundamental in-plane mechanical mode of the central nanobeam is approximately equivalent to a reduction in the width of one slot by -$x_{3}$ and an increase in the other slot by +$x_{3}$, the linear optomechanical coupling coefficient between optical slot mode $a_{1}$ and mechanical mode $b_{3}$ is estimated to be $\glin_{a_1,b_3} \approx (\glin_{a_1,\delta s_1}+\glin_{a_1,-\delta s_2}) = (\glin_{a_1,\delta s_1}-\glin_{a_2,\delta s_1}) = 2\pi[136$~GHz/nm$]$, where by symmetry $\glin_{a_1,-\delta s_2} = -\glin_{a_2,\delta s_1}$.  Along with a measured splitting of $2J/2\pi = 12$~GHz, this yields through eq.~(\ref{eq:g'}) an estimate for the $x^2$-coupling coefficient of $\gsq_{b_3}/2\pi \approx 1.54$~THz/nm$^2$.  



%
%
%


\subsection{Transduction of mechanical motion}
\label{subsec:suppresion_lin}

\begin{figure}[!t]
\begin{center}
\includegraphics[width=\columnwidth]{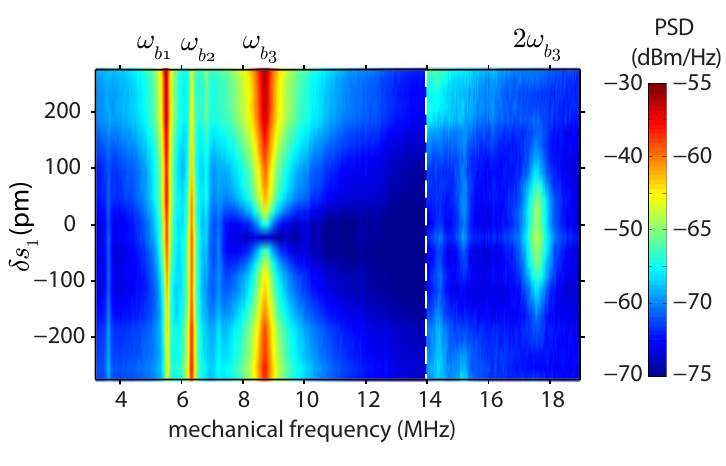}
\caption{RF photocurrent noise spectrum for the optically transmitted light past the double-slotted photonic crystal cavity.  Here the applied voltage $V_{2}=1$~V is held fixed and $V_{1}$ is swept from just below to just above the anti-crossing point of Fig.~\ref{fig_tuning_small_slot}a.  In these measurements the probe laser power is $10$~$\mu$W at the input to the cavity, the probe laser frequency is set on the blue side of the upper frequency supermode resonance, $\Delta_L \approx \kappa/2\sqrt{3}$, and the fiber taper is placed in the near-field of the photonic crystal cavity resulting in an on-resonance dip in transmission of approximately $\Delta T = 15$\%.  The vertical axis in this plot is converted to a change in the slot width, $\delta s_{1}$, using the numerically simulated value of $\alpha_{\text{cap}}=0.025$~nm/V$^{2}$.  The color indicates the magnitude of the RF noise in dBm/Hz, where the colorscale from $0$-$14$~MHz is shown on the left of the scalebar and the colorscale from $14$-$20$~MHz is shown on the right of the scalebar (a different scale is used to highlight the noise out at $2\omega_{b_{3}}$.}\label{fig_trapping}
\end{center}
\end{figure}

Figure.~\ref{fig_trapping} shows the evolution of the optically-transduced mechanical noise power spectral density (PSD) near the anti-crossing region of Fig.~\ref{fig_tuning_small_slot}a.  In this plot $s_{2}$ is fixed and $s_{1}$ is varied over an estimated range of $\delta s_{1}=\pm 0.3$~nm around the anti-crossing.  Optical motion is imprinted as intensity modulations of the probe laser which is tuned to the blue side of the upper frequency supermode.  Here we choose the detuning point corresponding to $\Delta_L \equiv \omega_L - \omega_{+} \approx \kappa/2\sqrt{3}$, where $\omega_{L}$ is the probe laser frequency and $\kappa$ is the full-width at half-maximum linewidth of the optical resonance.  This detuning choice ensures (maximal) linear transduction of small fluctuations in the frequency of the cavity supermode, which allows us to relate nonlinear transduction of motion with true nonlinear optomechanical coupling~\cite{PhysRevA.89.053838,Kaviani2014-paddle}.  A probe power of $\Pin = 10$~$\mu$W is used in order to avoid any nonlinear effects due to optical absorption, and the transmitted light is first amplified through an erbium-doped fiber amplifier before being detected on a high gain photoreceiver (transimpedance gain $10^4$~V/A, NEP$=12$~pW/Hz$^{1/2}$, bandwidth $150$~MHz).  The resulting radio-frequency (RF) photocurrent noise spectrum is plotted in Fig.~\ref{fig_trapping}.

To help identify the measured noise peaks numerical FEM simulations of the mechanical properties of the double-slotted structure were performed.  Taking structural dimensions from SEM images, the simulated mechanical frequency for the fundamental in-plane resonances of the two outer slabs ($b_{1}$ and $b_{2}$) is found to be $\omegam/2\pi = 8.4$MHz. An effective motional mass for the slab modes of $\meff = 35$~pg was determined by integrating, over the volume of the structure, the mass density of the silicon slab weighted by the normalized, squared displacement amplitude of the slab's motion~\cite{Eichenfield:09}.  The corresponding estimate of the zero-point amplitude of the slab modes is given by $\xzpf \equiv (\hbar/2\meff\omegam)^{1/2} = 5.6$~fm. The resonance frequency, effective motional mass, and zero-point amplitude for the fundamental in-plane resonance of the central nanobeam ($b_{3}$) are simulated to be $\omegam/2\pi = 10.7$MHz, $m = 3.6$~pg, and $\xzpf = 15.4$~fm, respectively.

Comparing to Fig.~\ref{fig_trapping}a, the two lowest frequency noise peaks are thus identified as due to the thermal motion of the $b_1$ and $b_2$ modes of the outer slabs, with $\omega_{b_1}/2\pi=5.54$~MHz and $\omega_{b_2}/2\pi=6.34$~MHz.  The identification of the $b_{1}$ mode with the lower frequency mechanical resonance is made possible due to the increasing signal transduction of this resonance as $s_{1}$ is increased above the anti-crossing point.  Since we are probing the upper frequency optical supermode, for $s_{1} > s_{2}$ ($\delta s_{1} > 0$) the supermode is approximately $a_{1}$ which is localized to slot 1 and sensitive primarily to the motion of $b_{1}$.  We see an opposite trend for the $b_{2}$ resonance, with larger transduction gain for $s_{1} < s_{2}$ ($\delta s_{1} < 0$).  The frequencies of both these modes is lower than found in numerical simulations, likely due to squeeze-film damping effects not captured in the FEM analysis~\cite{Bao2007}.  

The noise peak at $\omegam/2\pi = 8.73$~MHz behaves altogether differently than the $b_{1}$ and $b_{2}$ resonances, and is identified with the $b_{3}$ mode of the central nanobeam (although again at a lower frequency than expected from FEM simulation).  This noise peak is transduced with roughly equal signal levels for $\delta s_{1} > 0$ and $\delta s_{1} < 0$, but significantly drops in strength for $\delta s_{1} \approx 0$ near the anti-crossing.  This is the expected characteristic of the $b_{3}$ mode, where the dispersive linear optomechanical coupling to the $b_{3}$ should vanish at the anti-crossing point.  Also shown in Fig.~\ref{fig_trapping}a is the noise at $2\omega_{b_{3}}/2\pi \approx 17.5$~MHz, which shows a weakly transduced resonance with signal strength peaked around $\delta s_{1} = 0$.  The suppression in transduction of the noise peak at $\omega_{b_{3}}$ concurrent with the rise in transduction of the noise peak at $2\omega_{b_{3}}$ is a direct manifestation of the transition from linear ($\glin_{a_{1},b_{3}}$ or $\glin_{a_{2},b_{3}}$) to position-squared ($\gsq_{b_{3}}$) optomechanical coupling.  

\subsection{Static and dynamic optical spring measurements}
\label{subsec:spring_effect}

Our previous estimate of $\gsq_{b_{3}}$ from the anti-crossing curves relied on the approximate correspondence between the static displacement of the outer slabs and the fundamental in-plane vibrational amplitude of the $b_{3}$ mode of the central nanobeam.  A more accurate determination of the true $x^2$-coupling coefficient to $b_{3}$ can be determined from two different optical spring measurements.  Far from the anti-crossing one can determine the linear optomechanical coupling coefficient between the optical slot modes and the $b_{3}$ mechanical mode from the dynamic back-action of the intra-cavity light field on the mechanical frequency, which in conjunction with the measured anti-crossing splitting yields $\gsq_{b_{3}}$ via eq.~(\ref{eq:g'}).  A direct measurement of $\gsqzpf_{b_{3}}$ can also be obtained from the static optical spring effect near the anti-crossing point as indicated in eq.~(\ref{eq:HRWA}).
 
\begin{figure}[!t]
\begin{center}
\includegraphics[width=\columnwidth]{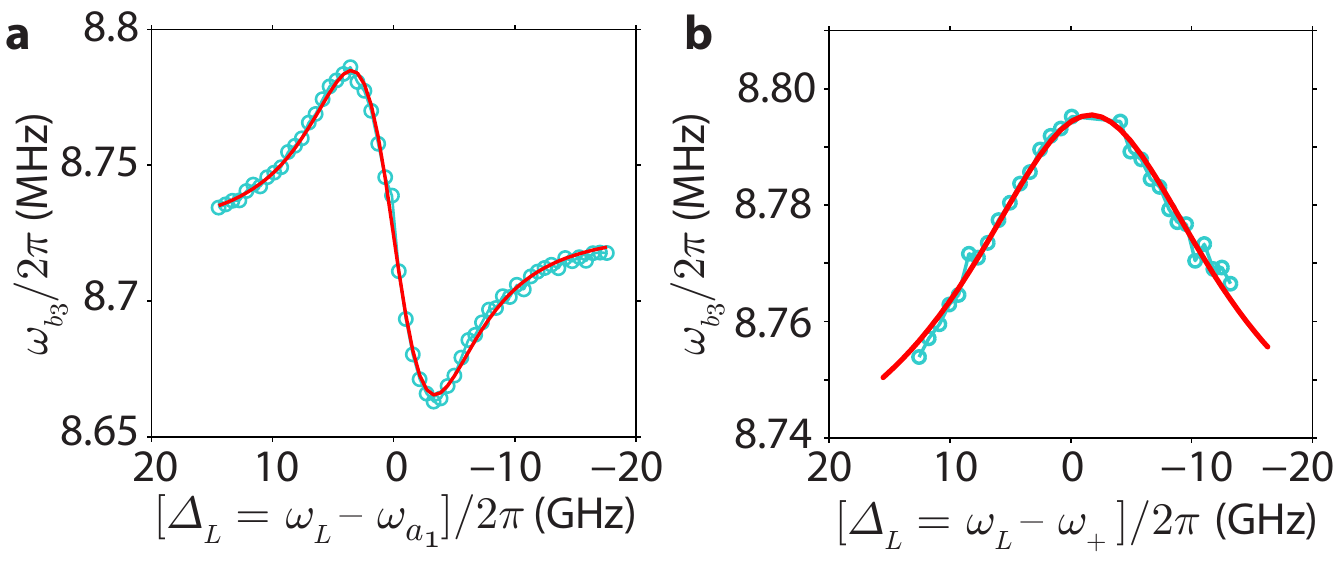} 
\caption{(a) Dynamic optical spring effect measured by exciting the upper frequency supermode resonance far from the anti-crossing point ($\sim a_{1}$ mode) [$V_{1}=V_{2}=1$~V, $\Pin = 10$~$\mu$W, $\kappa/2\pi=12.5$~GHz, $\Delta T \approx 10\%$]. (b) Static optical spring shift of the $b_{3}$ resonance frequency versus laser detuning $\Delta_{L}$ from the upper ($\sim a_{+}$) supermode resonance near the anti-crossing point [$V_{1}=10.9$~V, $V_{2}=1$~V, $\Pin = 50$~$\mu$W, $\kappa/2\pi=26$~GHz, $\Delta T \approx 25\%$].  In both (a) and (b) $V_{2}$ is fixed at $1$~V (see Fig.~\ref{fig_tuning_small_slot}a) and the measured data (circles) correspond to a Lorentzian fit to the resonance freqency of the optically transduced thermal noise peak at $\omega_{b_3}$.  In (a) the red curve is a fit to the data using a dynamical optical spring model~\cite{Eichenfield:09} with linear optomechanical coupling coefficient $\tilde{\glin}_{a_{1},b_{3}}/2\pi=1.72$~MHz.  In (b) the red curve is a fit to the data using a static spring model (c.f., eqs.~(\ref{eq:H_AS_QS_spring}) and (\ref{eq:HRWA})) with $x^2$-coupling coefficient $\tilde{g}^{\prime}_{b_3}/2\pi=46$~Hz.  In both the spring models of (a) and (b) the intra-cavity photon number versus detuning $n(\Delta_{L})$ is calibrated from the known input laser power, cavity linewidth, and on-resonance transmission contrast.  
\label{fig_spring}}
\end{center}
\end{figure}

Figure~\ref{fig_spring}a shows the dependence of the mechanical resonance frequency of the $b_{3}$ mode of the central nanobeam versus the laser detuning $\Delta_L$ when the device is tuned far from the anti-crossing point in Fig.~\ref{fig_tuning_small_slot}a ($V_{1}= 1$~V and $V_{2}=1$~V).  In these measurements the probe laser power is fixed at $\Pin = 10$~$\mu$W and the laser frequency is scanned across the upper optical supermode resonance, which away from the anti-crossing point in this case is the slot-mode $a_{1}$.  In the sideband unresolved regime ($\omegam \ll \kappa$), the dynamic optical spring effect has a dispersive lineshape centered around the optical resonance frequency, with optical softening of the mechanical resonance occurring for red detuning ($\Delta_L < 0$) and optical stiffening occurring for blue detuning ($\Delta_L >0$).  

A fit to the measured frequency shift versus $\Delta_L$ is performed using the linear optomechanical coupling rate $\tilde{\glin}_{a_{1},b_{3}}$ as a fit parameter.  The resulting optomechanical coupling rate which best fits the data is shown in Figure~\ref{fig_spring}a as a red curve, and corresponds to $\tilde{\glin}_{a_{1},b_{3}}/2\pi = 1.72$~MHz.  Using $\xzpf=16$~fm for the $b_{3}$ mechanical mode, this corresponds to $\glin_{a_{1},b_{3}}/2\pi = 107$~GHz/nm.  Note that this is slightly smaller than the value measured indirectly from the dispersion in the anti-crossing curve of Fig.~\ref{fig_tuning_small_slot}, however, that value relied on the simulated value for $\alphacap$ which is quite sensitive to the actual fabricated dimensions and stiffness of the structure. For the smallest splitting measured in this work ($2J/2\pi = 12$~GHz), we get an estimated value for the $x^2$-coupling to the $b_{3}$ mode from the dynamic optical spring measurements of $\gsqzpf_{b_3}/2\pi = 245$~Hz ($\gsq_{b_3}/2\pi=0.96$~THz/nm$^2$).         



An entirely different dynamics occurs at the anti-crossing point where $x^2$ optomechanical coupling dominates.  Optical pumping of the supermode resonances near the anti-crossing point gives rise to an optical spring shift which depends on the static (i.e., not how it modulates with motion) value of the intracavity photon number.  Due to the opposite sign of the quadratic dispersion of the upper and lower optical supermode branches, optical pumping of the upper branch resonance leads to a stiffening of the mechanical structure whereas optical pumping of the lower branch leads to a softening of the structure~\cite{PhysRevA.89.053838,Lee2015-x2spring}.  The measured frequency shift of the $b_{3}$ mechanical resonance for optical pumping of the upper branch cavity supermode (the even $a_{+}$ mode in this case) is shown in Fig.~\ref{fig_spring}b for a voltage setting on the capacitor electrodes of $V_{1}=10.6$~V and $V_{2}=1$~V.  This position is slightly below the exact center of the anti-crossing point of Fig.~\ref{fig_tuning_small_slot}a so as to allow weak linear transduction of the $b_{3}$ resonance.  A rather large supermode splitting of $2J/2\pi = 50$~GHz was also chosen to ensure that only the even $a_{+}$ supermode is excited, and that the contribution to the optical trapping (anti-trapping) by the lower branch $a_{-}$ resonance is negligible.  

As per eqs.~(\ref{eq:H_AS_QS_spring}) and (\ref{eq:HRWA}), the mechanical frequency shift is approximately given by $\Delta\omega_m(\Delta_L) \approx 2 \gsqzpf_{b_{3}} n_{+}(\Delta_L)$, where $n_{+}(\Delta_L)$ is the average intra-cavity photon number in the $a_{+}$ supermode.  Fitting this model to the data measured in Fig.~\ref{fig_spring}b yields a value of $\gsqzpf_{b_3}/2\pi=46$~Hz.  This is slightly lower than the $60$~Hz value expected for a splitting of $2J/2\pi = 50$~GHz and the linear coupling rate of $\tilde{\glin}_{a_{1},b_{3}}/2\pi = 1.72$~MHz determined from the dynamical optical spring effect, but consistent with our slight detuning of the structure from the exact center of the anti-crossing.           


\section{Discussion}
\label{sec:disc}

The quasi two-dimensional photonic crystal architecture as presented here provides a means of realizing extremely large dispersive $x^2$-coupling between light and mechanics.  This is due to the ability to co-localize optical and acoustic waves in a common wavelength scale volume, resulting in inherently large linear optomechanical coupling.  Combined with an ability to engineer the optical mode dispersion to allow for a tunable degree of optical mode splitting, the $x^2$-coupling can be even further enhanced.  It is interesting to consider then, just how far this technology could be pushed given recent technical advances made in the area of photonic crystals and optomechanical crystals.

We consider here the feasibility of a QND measurement of phonon number, although similar parameters would enable a measurement of phonon shot noise~\cite{Clerk_Phonon_Shot_Noise}, a QND measurement of photon number~\cite{Ludwig_enhanced}, and the cooling and squeezing of mechanical motion~\cite{PhysRevA.77.033819,PhysRevA.82.021806,PhysRevA.89.023849}. In the quasi-static limit as realized in this work ($|J| \gg \omegam$), the optical resonance shift per phonon is $\Delta\omega = 2\gsqzpf = \glinzpf^2/J$.  If the lower frequency optical resonance ($a_{-}$ in the case $J>0$) is used to probe the system, then roughly the photons emitted per unit time from the $a_{-}$ cavity mode would change by $(n_{-}\kappa_{-})(\Delta\omega/\kappa_{-})$ upon a single phonon jump in the mechanical resonator.  Assuming shot-noise limited detection over a measurement time $\tau$, the signal to noise ratio (SNR) for a phonon jump is given approximately by,

\begin{equation}\label{eq:SNR}
\SNR \approx \frac{(n_{-}\kappa_{-})^2(\Delta\omega/\kappa_{-})^2\tau^2}{n_{-}\kappa_{-}\tau} = \left[\frac{n_{-}\Delta\omega^2}{\kappa_{-}}\right]\tau. 
\end{equation}         

\noindent The corresponding phonon jump measurement rate follows from the term in the bracket of eq.~(\ref{eq:SNR}), $\Gammameas = \left[(\Delta\omega)^2/\kappa_{-}\right]n_{-} = \left[4(\gsqzpf)^2/\kappa_{-}\right]n_{-}$.  

This measurement rate should be compared against the decoherence rate of the mechanical resonator.  The thermal decoherence rate is $\Gammath = (\nbath + 1)\gammami$, where $\nbath$ is the Bose occupation factor depending on the bath temperature ($\Tbath$), and $\gammami$ is the intrinsic mechanical damping rate to the bath.  At $\Tbath = 4$~K similar silicon photonic crystal devices have been operated with intra-cavity photon numbers of $10^3$ and mechanical $Q$-factor as large as $7\times 10^5$~\cite{Safavi-Naeini2013-squeezing}.  For the device studied here ($\omegam/2\pi \approx 10$~MHz, $\gsqzpf/2\pi = 240$~Hz, $\kappa_{-}/2\pi = 5$~GHz), the phonon jump measurement rate would be $\Gammameas/2\pi \approx 46$~mHz, while the thermal decoherence rate at $\Tbath=4$~K and for $\Qm = 7\times10^5$ is $\Gammath/2\pi \approx 125$~kHz.  Significant improvements in the measurement rate can be realized with improved optical $Q$-factor.  Recent work by Sekoguchi, et al.~\cite{Sekoguchi2014-9millionQ}, has shown that optical $Q$-factors of order $10^{7}$ can be realized in similar planar 2D silicon photonic crystals in the telecom band, corresponding to a minimum cavity decay rate of $\kappa/2\pi = 20$~MHz.  By proper tuning of the double-slotted photonic crystal structure, the optical mode splitting $2J$ could be reduced down to a minimum resolvable value equal to $\kappa$, yielding an $x^2$-coupling value of $\gsqzpf/2\pi \approx 100$~kHz and a phonon jump measurement rate of $\Gammameas/2\pi \approx 2$~MHz.

In order to realize a sideband-resolved system, higher mechanical resonant frequencies must also be employed.  Numerical simulations~\cite{Kalaee2015} indicate that higher-order modes of the central nanobeam can maintain significant optomechanical coupling, with $\glinzpf/2\pi \approx 0.4$~MHz for the seventh-order in-plane mechanical resonance at $\omegam/2\pi = 225$~MHz.  Tuning the structure such that the mode splitting is nearly resonant with the mechanical frequency, $\glinzpf \ll |\delta \equiv |2J| - \omegam| \ll \omegam, |2J|$, greatly enhances the frequency shift per phonon as per eq.~(\ref{eq:Heffres}), $\Delta\omega=\glinzpf^2/\delta$.  For similar cavity conditions as above ($n_{-} = 10^3$, $\kappa_{-}/2\pi = 20$~MHz), and assuming $\delta = 10\glinzpf$, yields a measurement rate of $\Gammameas \approx 80$~kHz.  This is comparable to the thermal decoherence rate at $\Tbath=4$~K assuming a similar mechanical $Q$-factor for these higher frequency modes.  Recent measurements at bath temperatures of $\Tbath \lesssim 100$~mK, however, have shown that mechanical $Q$-factors in excess of $10^7$ can be realized in silicon using phononic bandgap acoustic shielding patterned in the perimeter of the device~\cite{Meenehan2014}. At these temperatures we can expect a bath occupancy of $\nbath \approx 10$, and with an acoustic bandgap shield, a much smaller thermal decoherence rate of $\Gammath/2\pi \approx 300$~Hz.  A comparable measurement rate could then be employed with a much weaker optical probe corresponding to an intra-cavity photon number of $n_{-} \approx 10$.               

The most challenging aspect of a QND phonon number measurement, however, is the optically induced mechanical decay due to residual back-action stemming from the linear (in $\hat{x}$) cross-coupling of the cavity supermodes~\cite{Ludwig_enhanced,Miao_standard_quantum_limit}.  This parasitic back-action damping of the mechanical resonator occurs through a process, for example, in which a photon is scattered from the driven $a_{-}$ mode into the $a_{+}$ mode where it decays into the optical bath, absorbing a phonon in the process.  The optically-induced mechanical decay rate for a $n_{b}$-phonon Fock state is given by $\Gammaopt \approx \left(\glinzpf/\delta\right)^2n_{b} n_{-}\kappa_{-}$ ($=\left((\gsqzpf)^2/|2J|\right)n_{b} n_{-}\kappa_{+}$ in the quasi-static limit)~\cite{Ludwig_enhanced}. Comparing to the phonon jump measurement rate we see that only in the vacuum strong coupling limit ($\glinzpf/\kappa \gtrsim 1$) can one realize a continuous QND measurement of phonon number,

\begin{equation}\label{eq:}
\frac{\Gammameas}{\Gammaopt} \approx \frac{\glinzpf^2}{n_{b}\kappa_{+}\kappa_{-}} \lesssim \left(\frac{\glinzpf}{\kappa}\right)^2. 
\end{equation}

\noindent Note that a more careful analysis~\cite{Ludwig_enhanced,Miao_standard_quantum_limit} indicates that a limit of $\glinzpf \gtrsim \kappa_{i}$ need only be met, where $\kappa_{i}$ is the intrinsic damping of the optical cavity excluding loading of the cavity by measurement channels.  A ratio of $\glinzpf/\kappa_{i} \approx 0.007$ has previously been realized in silicon optomechanical crystals~\cite{Chan2012}. In the case of the double-slotted photonic crystal structure studied here, fabrication of nanoscale slots as small as $s=25$~nm~\cite{Pitanti2015} would increase the linear optomechanical coupling between $a_{\pm}$ cavity supermodes to $\glinzpf/2\pi \sim 10$~MHz.  With this advance, and in conjunction with an increase of the optical $Q$-factor to $10^7$~\cite{Sekoguchi2014-9millionQ}, it does seem feasible in the near future to reach the vacuum strong coupling limit which would enable QND phononic and photonic measurements as proposed in Ref.~\cite{Ludwig_enhanced}.

\begin{acknowledgments}
The authors would like to thank M. Davanco and Max Ludwig for fruitful discussions. This work was supported by the AFOSR Hybrid Nanophotonics MURI, the Institute for Quantum Information and Matter, an NSF Physics Frontiers Center with support of the Gordon and Betty Moore Foundation, the Alexander von Humboldt Foundation, the Max Planck Society, and the Kavli Nanoscience Institute at Caltech. FM acknowledges support from the DARPA ORCHID program, ERC OPTOMECH, and ITN cQOM.  TKP gratefully acknowledges support from the Swiss National Science Foundation. MK was supported by a College Doctoral Fellowship from the University of Southern California.
\end{acknowledgments}

\bibliographystyle{apsrev4-1}

\begin{thebibliography}{45}%
\makeatletter
\providecommand \@ifxundefined [1]{%
 \@ifx{#1\undefined}
}%
\providecommand \@ifnum [1]{%
 \ifnum #1\expandafter \@firstoftwo
 \else \expandafter \@secondoftwo
 \fi
}%
\providecommand \@ifx [1]{%
 \ifx #1\expandafter \@firstoftwo
 \else \expandafter \@secondoftwo
 \fi
}%
\providecommand \natexlab [1]{#1}%
\providecommand \enquote  [1]{``#1''}%
\providecommand \bibnamefont  [1]{#1}%
\providecommand \bibfnamefont [1]{#1}%
\providecommand \citenamefont [1]{#1}%
\providecommand \href@noop [0]{\@secondoftwo}%
\providecommand \href [0]{\begingroup \@sanitize@url \@href}%
\providecommand \@href[1]{\@@startlink{#1}\@@href}%
\providecommand \@@href[1]{\endgroup#1\@@endlink}%
\providecommand \@sanitize@url [0]{\catcode `\\12\catcode `\$12\catcode
  `\&12\catcode `\#12\catcode `\^12\catcode `\_12\catcode `\%12\relax}%
\providecommand \@@startlink[1]{}%
\providecommand \@@endlink[0]{}%
\providecommand \url  [0]{\begingroup\@sanitize@url \@url }%
\providecommand \@url [1]{\endgroup\@href {#1}{\urlprefix }}%
\providecommand \urlprefix  [0]{URL }%
\providecommand \Eprint [0]{\href }%
\providecommand \doibase [0]{http://dx.doi.org/}%
\providecommand \selectlanguage [0]{\@gobble}%
\providecommand \bibinfo  [0]{\@secondoftwo}%
\providecommand \bibfield  [0]{\@secondoftwo}%
\providecommand \translation [1]{[#1]}%
\providecommand \BibitemOpen [0]{}%
\providecommand \bibitemStop [0]{}%
\providecommand \bibitemNoStop [0]{.\EOS\space}%
\providecommand \EOS [0]{\spacefactor3000\relax}%
\providecommand \BibitemShut  [1]{\csname bibitem#1\endcsname}%
\let\auto@bib@innerbib\@empty
\bibitem [{\citenamefont {Aspelmeyer}\ \emph {et~al.}(2014)\citenamefont
  {Aspelmeyer}, \citenamefont {Kippenberg},\ and\ \citenamefont
  {Marquardt}}]{RevModPhys.86.1391}%
  \BibitemOpen
  \bibfield  {author} {\bibinfo {author} {\bibfnamefont {M.}~\bibnamefont
  {Aspelmeyer}}, \bibinfo {author} {\bibfnamefont {T.~J.}\ \bibnamefont
  {Kippenberg}}, \ and\ \bibinfo {author} {\bibfnamefont {F.}~\bibnamefont
  {Marquardt}},\ }\href {\doibase 10.1103/RevModPhys.86.1391} {\bibfield
  {journal} {\bibinfo  {journal} {Rev. Mod. Phys.}\ }\textbf {\bibinfo {volume}
  {86}},\ \bibinfo {pages} {1391} (\bibinfo {year} {2014})}\BibitemShut
  {NoStop}%
\bibitem [{\citenamefont {Purdy}\ \emph {et~al.}(2010)\citenamefont {Purdy},
  \citenamefont {Brooks}, \citenamefont {Botter}, \citenamefont {Brahms},
  \citenamefont {Ma},\ and\ \citenamefont
  {Stamper-Kurn}}]{PhysRevLett.105.133602}%
  \BibitemOpen
  \bibfield  {author} {\bibinfo {author} {\bibfnamefont {T.~P.}\ \bibnamefont
  {Purdy}}, \bibinfo {author} {\bibfnamefont {D.~W.~C.}\ \bibnamefont
  {Brooks}}, \bibinfo {author} {\bibfnamefont {T.}~\bibnamefont {Botter}},
  \bibinfo {author} {\bibfnamefont {N.}~\bibnamefont {Brahms}}, \bibinfo
  {author} {\bibfnamefont {Z.-Y.}\ \bibnamefont {Ma}}, \ and\ \bibinfo {author}
  {\bibfnamefont {D.~M.}\ \bibnamefont {Stamper-Kurn}},\ }\href {\doibase
  10.1103/PhysRevLett.105.133602} {\bibfield  {journal} {\bibinfo  {journal}
  {Phys. Rev. Lett.}\ }\textbf {\bibinfo {volume} {105}},\ \bibinfo {pages}
  {133602} (\bibinfo {year} {2010})}\BibitemShut {NoStop}%
\bibitem [{\citenamefont {Eichenfield}\ \emph
  {et~al.}(2009{\natexlab{a}})\citenamefont {Eichenfield}, \citenamefont
  {Chan}, \citenamefont {Camacho}, \citenamefont {Vahala},\ and\ \citenamefont
  {Painter}}]{mattopto2009}%
  \BibitemOpen
  \bibfield  {author} {\bibinfo {author} {\bibfnamefont {M.}~\bibnamefont
  {Eichenfield}}, \bibinfo {author} {\bibfnamefont {J.}~\bibnamefont {Chan}},
  \bibinfo {author} {\bibfnamefont {R.}~\bibnamefont {Camacho}}, \bibinfo
  {author} {\bibfnamefont {K.}~\bibnamefont {Vahala}}, \ and\ \bibinfo {author}
  {\bibfnamefont {O.}~\bibnamefont {Painter}},\ }\href {\doibase
  10.1038/nature08524} {\bibfield  {journal} {\bibinfo  {journal} {Nature}\
  }\textbf {\bibinfo {volume} {462}},\ \bibinfo {pages} {78} (\bibinfo {year}
  {2009}{\natexlab{a}})}\BibitemShut {NoStop}%
\bibitem [{\citenamefont {\textnormal{B. Abbott \emph{et al.} (LIGO Scientific
  Collaboration)}}(2009)}]{LIGO-near-ground-state}%
  \BibitemOpen
  \bibfield  {author} {\bibinfo {author} {\bibnamefont {\textnormal{B. Abbott
  \emph{et al.} (LIGO Scientific Collaboration)}}},\ }\href
  {http://stacks.iop.org/1367-2630/11/i=7/a=073032} {\bibfield  {journal}
  {\bibinfo  {journal} {New J. Phys.}\ }\textbf {\bibinfo {volume} {11}},\
  \bibinfo {pages} {073032} (\bibinfo {year} {2009})}\BibitemShut {NoStop}%
\bibitem [{\citenamefont {Safavi-Naeini}\ \emph {et~al.}(2011)\citenamefont
  {Safavi-Naeini}, \citenamefont {{Mayer Alegre}}, \citenamefont {Chan},
  \citenamefont {Eichenfield}, \citenamefont {Winger}, \citenamefont {Lin},
  \citenamefont {Hill}, \citenamefont {Chang},\ and\ \citenamefont
  {Painter}}]{Safavi-Naeini2011-EIT}%
  \BibitemOpen
  \bibfield  {author} {\bibinfo {author} {\bibfnamefont {A.~H.}\ \bibnamefont
  {Safavi-Naeini}}, \bibinfo {author} {\bibfnamefont {T.~P.}\ \bibnamefont
  {{Mayer Alegre}}}, \bibinfo {author} {\bibfnamefont {J.}~\bibnamefont
  {Chan}}, \bibinfo {author} {\bibfnamefont {M.}~\bibnamefont {Eichenfield}},
  \bibinfo {author} {\bibfnamefont {M.}~\bibnamefont {Winger}}, \bibinfo
  {author} {\bibfnamefont {Q.}~\bibnamefont {Lin}}, \bibinfo {author}
  {\bibfnamefont {J.~T.}\ \bibnamefont {Hill}}, \bibinfo {author}
  {\bibfnamefont {D.~E.}\ \bibnamefont {Chang}}, \ and\ \bibinfo {author}
  {\bibfnamefont {O.}~\bibnamefont {Painter}},\ }\href {\doibase
  10.1038/nature09933} {\bibfield  {journal} {\bibinfo  {journal} {Nature}\
  }\textbf {\bibinfo {volume} {472}},\ \bibinfo {pages} {69} (\bibinfo {year}
  {2011})}\BibitemShut {NoStop}%
\bibitem [{\citenamefont {Weis}\ \emph {et~al.}(2010)\citenamefont {Weis},
  \citenamefont {Rivi\`{e}re}, \citenamefont {Del\'{e}glise}, \citenamefont
  {Gavartin}, \citenamefont {Arcizet}, \citenamefont {Schliesser},\ and\
  \citenamefont {Kippenberg}}]{Weis2010-OMIT}%
  \BibitemOpen
  \bibfield  {author} {\bibinfo {author} {\bibfnamefont {S.}~\bibnamefont
  {Weis}}, \bibinfo {author} {\bibfnamefont {R.}~\bibnamefont {Rivi\`{e}re}},
  \bibinfo {author} {\bibfnamefont {S.}~\bibnamefont {Del\'{e}glise}}, \bibinfo
  {author} {\bibfnamefont {E.}~\bibnamefont {Gavartin}}, \bibinfo {author}
  {\bibfnamefont {O.}~\bibnamefont {Arcizet}}, \bibinfo {author} {\bibfnamefont
  {A.}~\bibnamefont {Schliesser}}, \ and\ \bibinfo {author} {\bibfnamefont
  {T.~J.}\ \bibnamefont {Kippenberg}},\ }\href {\doibase
  10.1126/science.1195596} {\bibfield  {journal} {\bibinfo  {journal}
  {Science}\ }\textbf {\bibinfo {volume} {330}},\ \bibinfo {pages} {1520}
  (\bibinfo {year} {2010})}\BibitemShut {NoStop}%
\bibitem [{\citenamefont {O'Connell}\ \emph {et~al.}(2010)\citenamefont
  {O'Connell}, \citenamefont {Hofheinz}, \citenamefont {Ansmann}, \citenamefont
  {Bialczak}, \citenamefont {Lenander}, \citenamefont {Lucero}, \citenamefont
  {Neeley}, \citenamefont {Sank}, \citenamefont {Wang}, \citenamefont {Weides},
  \citenamefont {Wenner}, \citenamefont {Martinis},\ and\ \citenamefont
  {Cleland}}]{OConnell2010-ground-state}%
  \BibitemOpen
  \bibfield  {author} {\bibinfo {author} {\bibfnamefont {A.~D.}\ \bibnamefont
  {O'Connell}}, \bibinfo {author} {\bibfnamefont {M.}~\bibnamefont {Hofheinz}},
  \bibinfo {author} {\bibfnamefont {M.}~\bibnamefont {Ansmann}}, \bibinfo
  {author} {\bibfnamefont {R.~C.}\ \bibnamefont {Bialczak}}, \bibinfo {author}
  {\bibfnamefont {M.}~\bibnamefont {Lenander}}, \bibinfo {author}
  {\bibfnamefont {E.}~\bibnamefont {Lucero}}, \bibinfo {author} {\bibfnamefont
  {M.}~\bibnamefont {Neeley}}, \bibinfo {author} {\bibfnamefont
  {D.}~\bibnamefont {Sank}}, \bibinfo {author} {\bibfnamefont {H.}~\bibnamefont
  {Wang}}, \bibinfo {author} {\bibfnamefont {M.}~\bibnamefont {Weides}},
  \bibinfo {author} {\bibfnamefont {J.}~\bibnamefont {Wenner}}, \bibinfo
  {author} {\bibfnamefont {J.~M.}\ \bibnamefont {Martinis}}, \ and\ \bibinfo
  {author} {\bibfnamefont {A.~N.}\ \bibnamefont {Cleland}},\ }\href {\doibase
  10.1038/nature08967} {\bibfield  {journal} {\bibinfo  {journal} {Nature}\
  }\textbf {\bibinfo {volume} {464}},\ \bibinfo {pages} {697} (\bibinfo {year}
  {2010})}\BibitemShut {NoStop}%
\bibitem [{\citenamefont {Chan}\ \emph {et~al.}(2011)\citenamefont {Chan},
  \citenamefont {Alegre}, \citenamefont {Safavi-Naeini}, \citenamefont {Hill},
  \citenamefont {Krause}, \citenamefont {Gr\"{o}blacher}, \citenamefont
  {Aspelmeyer},\ and\ \citenamefont {Painter}}]{Chan2011-ground-state-cooling}%
  \BibitemOpen
  \bibfield  {author} {\bibinfo {author} {\bibfnamefont {J.}~\bibnamefont
  {Chan}}, \bibinfo {author} {\bibfnamefont {T.~P.~M.}\ \bibnamefont {Alegre}},
  \bibinfo {author} {\bibfnamefont {A.~H.}\ \bibnamefont {Safavi-Naeini}},
  \bibinfo {author} {\bibfnamefont {J.~T.}\ \bibnamefont {Hill}}, \bibinfo
  {author} {\bibfnamefont {A.}~\bibnamefont {Krause}}, \bibinfo {author}
  {\bibfnamefont {S.}~\bibnamefont {Gr\"{o}blacher}}, \bibinfo {author}
  {\bibfnamefont {M.}~\bibnamefont {Aspelmeyer}}, \ and\ \bibinfo {author}
  {\bibfnamefont {O.}~\bibnamefont {Painter}},\ }\href {\doibase
  10.1038/nature10461} {\bibfield  {journal} {\bibinfo  {journal} {Nature}\
  }\textbf {\bibinfo {volume} {478}},\ \bibinfo {pages} {89} (\bibinfo {year}
  {2011})}\BibitemShut {NoStop}%
\bibitem [{\citenamefont {Teufel}\ \emph {et~al.}(2011)\citenamefont {Teufel},
  \citenamefont {Donner}, \citenamefont {Li}, \citenamefont {Harlow},
  \citenamefont {Allman}, \citenamefont {Cicak}, \citenamefont {Sirois},
  \citenamefont {Whittaker}, \citenamefont {Lehnert},\ and\ \citenamefont
  {Simmonds}}]{Teufel2011-cooling}%
  \BibitemOpen
  \bibfield  {author} {\bibinfo {author} {\bibfnamefont {J.~D.}\ \bibnamefont
  {Teufel}}, \bibinfo {author} {\bibfnamefont {T.}~\bibnamefont {Donner}},
  \bibinfo {author} {\bibfnamefont {D.}~\bibnamefont {Li}}, \bibinfo {author}
  {\bibfnamefont {J.~W.}\ \bibnamefont {Harlow}}, \bibinfo {author}
  {\bibfnamefont {M.~S.}\ \bibnamefont {Allman}}, \bibinfo {author}
  {\bibfnamefont {K.}~\bibnamefont {Cicak}}, \bibinfo {author} {\bibfnamefont
  {A.~J.}\ \bibnamefont {Sirois}}, \bibinfo {author} {\bibfnamefont {J.~D.}\
  \bibnamefont {Whittaker}}, \bibinfo {author} {\bibfnamefont {K.~W.}\
  \bibnamefont {Lehnert}}, \ and\ \bibinfo {author} {\bibfnamefont {R.~W.}\
  \bibnamefont {Simmonds}},\ }\href {\doibase 10.1038/nature10261} {\bibfield
  {journal} {\bibinfo  {journal} {Nature}\ }\textbf {\bibinfo {volume} {475}},\
  \bibinfo {pages} {359} (\bibinfo {year} {2011})}\BibitemShut {NoStop}%
\bibitem [{\citenamefont {Brooks}\ \emph {et~al.}(2012)\citenamefont {Brooks},
  \citenamefont {Botter}, \citenamefont {Schreppler}, \citenamefont {Purdy},
  \citenamefont {Brahms},\ and\ \citenamefont
  {Stamper-Kurn}}]{Brooks2012-squeezing}%
  \BibitemOpen
  \bibfield  {author} {\bibinfo {author} {\bibfnamefont {D.~W.~C.}\
  \bibnamefont {Brooks}}, \bibinfo {author} {\bibfnamefont {T.}~\bibnamefont
  {Botter}}, \bibinfo {author} {\bibfnamefont {S.}~\bibnamefont {Schreppler}},
  \bibinfo {author} {\bibfnamefont {T.~P.}\ \bibnamefont {Purdy}}, \bibinfo
  {author} {\bibfnamefont {N.}~\bibnamefont {Brahms}}, \ and\ \bibinfo {author}
  {\bibfnamefont {D.~M.}\ \bibnamefont {Stamper-Kurn}},\ }\href {\doibase
  10.1038/nature11325} {\bibfield  {journal} {\bibinfo  {journal} {Nature}\
  }\textbf {\bibinfo {volume} {488}},\ \bibinfo {pages} {476} (\bibinfo {year}
  {2012})}\BibitemShut {NoStop}%
\bibitem [{\citenamefont {Safavi-Naeini}\ \emph {et~al.}(2013)\citenamefont
  {Safavi-Naeini}, \citenamefont {Gr\"{o}blacher}, \citenamefont {Hill},
  \citenamefont {Chan}, \citenamefont {Aspelmeyer},\ and\ \citenamefont
  {Painter}}]{Safavi-Naeini2013-squeezing}%
  \BibitemOpen
  \bibfield  {author} {\bibinfo {author} {\bibfnamefont {A.~H.}\ \bibnamefont
  {Safavi-Naeini}}, \bibinfo {author} {\bibfnamefont {S.}~\bibnamefont
  {Gr\"{o}blacher}}, \bibinfo {author} {\bibfnamefont {J.~T.}\ \bibnamefont
  {Hill}}, \bibinfo {author} {\bibfnamefont {J.}~\bibnamefont {Chan}}, \bibinfo
  {author} {\bibfnamefont {M.}~\bibnamefont {Aspelmeyer}}, \ and\ \bibinfo
  {author} {\bibfnamefont {O.}~\bibnamefont {Painter}},\ }\href
  {http://www.ncbi.nlm.nih.gov/pubmed/23925241} {\bibfield  {journal} {\bibinfo
   {journal} {Nature}\ }\textbf {\bibinfo {volume} {500}},\ \bibinfo {pages}
  {185} (\bibinfo {year} {2013})}\BibitemShut {NoStop}%
\bibitem [{\citenamefont {Thompson}\ \emph {et~al.}(2008)\citenamefont
  {Thompson}, \citenamefont {Zwickl}, \citenamefont {Jayich}, \citenamefont
  {Marquardt}, \citenamefont {Girvin},\ and\ \citenamefont
  {Harris}}]{Thompson_mim}%
  \BibitemOpen
  \bibfield  {author} {\bibinfo {author} {\bibfnamefont {J.~D.}\ \bibnamefont
  {Thompson}}, \bibinfo {author} {\bibfnamefont {B.~M.}\ \bibnamefont
  {Zwickl}}, \bibinfo {author} {\bibfnamefont {A.~M.}\ \bibnamefont {Jayich}},
  \bibinfo {author} {\bibfnamefont {F.}~\bibnamefont {Marquardt}}, \bibinfo
  {author} {\bibfnamefont {S.~M.}\ \bibnamefont {Girvin}}, \ and\ \bibinfo
  {author} {\bibfnamefont {J.~G.~E.}\ \bibnamefont {Harris}},\ }\href
  {http://dx.doi.org/10.1038/nature06715} {\bibfield  {journal} {\bibinfo
  {journal} {Nature}\ }\textbf {\bibinfo {volume} {452}},\ \bibinfo {pages}
  {72–75} (\bibinfo {year} {2008})}\BibitemShut {NoStop}%
\bibitem [{\citenamefont {Miao}\ \emph {et~al.}(2009)\citenamefont {Miao},
  \citenamefont {Danilishin}, \citenamefont {Corbitt},\ and\ \citenamefont
  {Chen}}]{Miao_standard_quantum_limit}%
  \BibitemOpen
  \bibfield  {author} {\bibinfo {author} {\bibfnamefont {H.}~\bibnamefont
  {Miao}}, \bibinfo {author} {\bibfnamefont {S.}~\bibnamefont {Danilishin}},
  \bibinfo {author} {\bibfnamefont {T.}~\bibnamefont {Corbitt}}, \ and\
  \bibinfo {author} {\bibfnamefont {Y.}~\bibnamefont {Chen}},\ }\href {\doibase
  10.1103/PhysRevLett.103.100402} {\bibfield  {journal} {\bibinfo  {journal}
  {Phys. Rev. Lett.}\ }\textbf {\bibinfo {volume} {103}},\ \bibinfo {pages}
  {100402} (\bibinfo {year} {2009})}\BibitemShut {NoStop}%
\bibitem [{\citenamefont {Ludwig}\ \emph {et~al.}(2012)\citenamefont {Ludwig},
  \citenamefont {Safavi-Naeini}, \citenamefont {Painter},\ and\ \citenamefont
  {Marquardt}}]{Ludwig_enhanced}%
  \BibitemOpen
  \bibfield  {author} {\bibinfo {author} {\bibfnamefont {M.}~\bibnamefont
  {Ludwig}}, \bibinfo {author} {\bibfnamefont {A.~H.}\ \bibnamefont
  {Safavi-Naeini}}, \bibinfo {author} {\bibfnamefont {O.}~\bibnamefont
  {Painter}}, \ and\ \bibinfo {author} {\bibfnamefont {F.}~\bibnamefont
  {Marquardt}},\ }\href {\doibase 10.1103/PhysRevLett.109.063601} {\bibfield
  {journal} {\bibinfo  {journal} {Phys. Rev. Lett.}\ }\textbf {\bibinfo
  {volume} {109}},\ \bibinfo {pages} {063601} (\bibinfo {year}
  {2012})}\BibitemShut {NoStop}%
\bibitem [{\citenamefont {Clerk}\ \emph {et~al.}(2010)\citenamefont {Clerk},
  \citenamefont {Marquardt},\ and\ \citenamefont
  {Harris}}]{Clerk_Phonon_Shot_Noise}%
  \BibitemOpen
  \bibfield  {author} {\bibinfo {author} {\bibfnamefont {A.~A.}\ \bibnamefont
  {Clerk}}, \bibinfo {author} {\bibfnamefont {F.}~\bibnamefont {Marquardt}}, \
  and\ \bibinfo {author} {\bibfnamefont {J.~G.~E.}\ \bibnamefont {Harris}},\
  }\href {\doibase 10.1103/PhysRevLett.104.213603} {\bibfield  {journal}
  {\bibinfo  {journal} {Phys. Rev. Lett.}\ }\textbf {\bibinfo {volume} {104}},\
  \bibinfo {pages} {213603} (\bibinfo {year} {2010})}\BibitemShut {NoStop}%
\bibitem [{\citenamefont {Bhattacharya}\ \emph {et~al.}(2008)\citenamefont
  {Bhattacharya}, \citenamefont {Uys},\ and\ \citenamefont
  {Meystre}}]{PhysRevA.77.033819}%
  \BibitemOpen
  \bibfield  {author} {\bibinfo {author} {\bibfnamefont {M.}~\bibnamefont
  {Bhattacharya}}, \bibinfo {author} {\bibfnamefont {H.}~\bibnamefont {Uys}}, \
  and\ \bibinfo {author} {\bibfnamefont {P.}~\bibnamefont {Meystre}},\ }\href
  {\doibase 10.1103/PhysRevA.77.033819} {\bibfield  {journal} {\bibinfo
  {journal} {Phys. Rev. A}\ }\textbf {\bibinfo {volume} {77}},\ \bibinfo
  {pages} {033819} (\bibinfo {year} {2008})}\BibitemShut {NoStop}%
\bibitem [{\citenamefont {Nunnenkamp}\ \emph {et~al.}(2010)\citenamefont
  {Nunnenkamp}, \citenamefont {B\o{}rkje}, \citenamefont {Harris},\ and\
  \citenamefont {Girvin}}]{PhysRevA.82.021806}%
  \BibitemOpen
  \bibfield  {author} {\bibinfo {author} {\bibfnamefont {A.}~\bibnamefont
  {Nunnenkamp}}, \bibinfo {author} {\bibfnamefont {K.}~\bibnamefont
  {B\o{}rkje}}, \bibinfo {author} {\bibfnamefont {J.~G.~E.}\ \bibnamefont
  {Harris}}, \ and\ \bibinfo {author} {\bibfnamefont {S.~M.}\ \bibnamefont
  {Girvin}},\ }\href {\doibase 10.1103/PhysRevA.82.021806} {\bibfield
  {journal} {\bibinfo  {journal} {Phys. Rev. A}\ }\textbf {\bibinfo {volume}
  {82}},\ \bibinfo {pages} {021806} (\bibinfo {year} {2010})}\BibitemShut
  {NoStop}%
\bibitem [{\citenamefont {Asjad}\ \emph {et~al.}(2014)\citenamefont {Asjad},
  \citenamefont {Agarwal}, \citenamefont {Kim}, \citenamefont {Tombesi},
  \citenamefont {Giuseppe},\ and\ \citenamefont {Vitali}}]{PhysRevA.89.023849}%
  \BibitemOpen
  \bibfield  {author} {\bibinfo {author} {\bibfnamefont {M.}~\bibnamefont
  {Asjad}}, \bibinfo {author} {\bibfnamefont {G.~S.}\ \bibnamefont {Agarwal}},
  \bibinfo {author} {\bibfnamefont {M.~S.}\ \bibnamefont {Kim}}, \bibinfo
  {author} {\bibfnamefont {P.}~\bibnamefont {Tombesi}}, \bibinfo {author}
  {\bibfnamefont {G.~D.}\ \bibnamefont {Giuseppe}}, \ and\ \bibinfo {author}
  {\bibfnamefont {D.}~\bibnamefont {Vitali}},\ }\href {\doibase
  10.1103/PhysRevA.89.023849} {\bibfield  {journal} {\bibinfo  {journal} {Phys.
  Rev. A}\ }\textbf {\bibinfo {volume} {89}},\ \bibinfo {pages} {023849}
  (\bibinfo {year} {2014})}\BibitemShut {NoStop}%
\bibitem [{\citenamefont {Vanner}(2011)}]{Vanner_prX_2011}%
  \BibitemOpen
  \bibfield  {author} {\bibinfo {author} {\bibfnamefont {M.~R.}\ \bibnamefont
  {Vanner}},\ }\href@noop {} {\bibfield  {journal} {\bibinfo  {journal} {Phys.
  Rev. X}\ }\textbf {\bibinfo {volume} {1}},\ \bibinfo {pages} {021011}
  (\bibinfo {year} {2011})}\BibitemShut {NoStop}%
\bibitem [{\citenamefont {Safavi-Naeini}(2013)}]{Amir_thesis2013}%
  \BibitemOpen
  \bibfield  {author} {\bibinfo {author} {\bibfnamefont {A.~H.}\ \bibnamefont
  {Safavi-Naeini}},\ }\emph {\bibinfo {title} {{Quantum Optomechanics with
  Silicon Nanostructures}}},\ \href@noop {} {Ph.D. thesis},\ \bibinfo  {school}
  {California Institute of Technology} (\bibinfo {year} {2013})\BibitemShut
  {NoStop}%
\bibitem [{\citenamefont {Kaviani}\ \emph {et~al.}(2014)\citenamefont
  {Kaviani}, \citenamefont {Healey}, \citenamefont {Wu}, \citenamefont
  {Ghobadi}, \citenamefont {Hryciw},\ and\ \citenamefont
  {Barclay}}]{Kaviani2014-paddle}%
  \BibitemOpen
  \bibfield  {author} {\bibinfo {author} {\bibfnamefont {H.}~\bibnamefont
  {Kaviani}}, \bibinfo {author} {\bibfnamefont {C.}~\bibnamefont {Healey}},
  \bibinfo {author} {\bibfnamefont {M.}~\bibnamefont {Wu}}, \bibinfo {author}
  {\bibfnamefont {R.}~\bibnamefont {Ghobadi}}, \bibinfo {author} {\bibfnamefont
  {A.}~\bibnamefont {Hryciw}}, \ and\ \bibinfo {author} {\bibfnamefont {P.~E.}\
  \bibnamefont {Barclay}},\ }\href@noop {} {\bibfield  {journal} {\bibinfo
  {journal} {Optica}\ }\textbf {\bibinfo {volume} {2}},\ \bibinfo {pages} {271}
  (\bibinfo {year} {2014})}\BibitemShut {NoStop}%
\bibitem [{\citenamefont {Hill}(2013)}]{Hill2013}%
  \BibitemOpen
  \bibfield  {author} {\bibinfo {author} {\bibfnamefont {J.~T.}\ \bibnamefont
  {Hill}},\ }\emph {\bibinfo {title} {{Nonlinear optics and wavelength
  translation via cavity-optomechanics}}},\ \href@noop {} {Ph.D. thesis},\
  \bibinfo  {school} {California Institute of Technology} (\bibinfo {year}
  {2013})\BibitemShut {NoStop}%
\bibitem [{\citenamefont {Doolin}\ \emph {et~al.}(2014)\citenamefont {Doolin},
  \citenamefont {Hauer}, \citenamefont {Kim}, \citenamefont {MacDonald},
  \citenamefont {Ramp},\ and\ \citenamefont {Davis}}]{PhysRevA.89.053838}%
  \BibitemOpen
  \bibfield  {author} {\bibinfo {author} {\bibfnamefont {C.}~\bibnamefont
  {Doolin}}, \bibinfo {author} {\bibfnamefont {B.~D.}\ \bibnamefont {Hauer}},
  \bibinfo {author} {\bibfnamefont {P.~H.}\ \bibnamefont {Kim}}, \bibinfo
  {author} {\bibfnamefont {A.~J.~R.}\ \bibnamefont {MacDonald}}, \bibinfo
  {author} {\bibfnamefont {H.}~\bibnamefont {Ramp}}, \ and\ \bibinfo {author}
  {\bibfnamefont {J.~P.}\ \bibnamefont {Davis}},\ }\href {\doibase
  10.1103/PhysRevA.89.053838} {\bibfield  {journal} {\bibinfo  {journal} {Phys.
  Rev. A}\ }\textbf {\bibinfo {volume} {89}},\ \bibinfo {pages} {053838}
  (\bibinfo {year} {2014})}\BibitemShut {NoStop}%
\bibitem [{\citenamefont {Brawley}\ \emph {et~al.}(2014)\citenamefont
  {Brawley}, \citenamefont {Vanner}, \citenamefont {Larsen}, \citenamefont
  {Schmid}, \citenamefont {Boisen},\ and\ \citenamefont
  {Bowen}}]{Brawley2014-nanostring}%
  \BibitemOpen
  \bibfield  {author} {\bibinfo {author} {\bibfnamefont {G.~A.}\ \bibnamefont
  {Brawley}}, \bibinfo {author} {\bibfnamefont {M.~R.}\ \bibnamefont {Vanner}},
  \bibinfo {author} {\bibfnamefont {P.~E.}\ \bibnamefont {Larsen}}, \bibinfo
  {author} {\bibfnamefont {S.}~\bibnamefont {Schmid}}, \bibinfo {author}
  {\bibfnamefont {A.}~\bibnamefont {Boisen}}, \ and\ \bibinfo {author}
  {\bibfnamefont {W.~P.}\ \bibnamefont {Bowen}},\ }\href
  {http://arxiv.org/abs/1404.5746} {\bibfield  {journal} {\bibinfo  {journal}
  {arXiv:1404.5746}\ } (\bibinfo {year} {2014})}\BibitemShut {NoStop}%
\bibitem [{\citenamefont {Flowers-Jacobs}\ \emph {et~al.}(2012)\citenamefont
  {Flowers-Jacobs}, \citenamefont {Hoch}, \citenamefont {Sankey}, \citenamefont
  {Kashkanova}, \citenamefont {Jayich}, \citenamefont {Deutsch}, \citenamefont
  {Reichel},\ and\ \citenamefont {Harris}}]{Flowers-jacobs_fiber}%
  \BibitemOpen
  \bibfield  {author} {\bibinfo {author} {\bibfnamefont {N.~E.}\ \bibnamefont
  {Flowers-Jacobs}}, \bibinfo {author} {\bibfnamefont {S.~W.}\ \bibnamefont
  {Hoch}}, \bibinfo {author} {\bibfnamefont {J.~C.}\ \bibnamefont {Sankey}},
  \bibinfo {author} {\bibfnamefont {A.}~\bibnamefont {Kashkanova}}, \bibinfo
  {author} {\bibfnamefont {A.~M.}\ \bibnamefont {Jayich}}, \bibinfo {author}
  {\bibfnamefont {C.}~\bibnamefont {Deutsch}}, \bibinfo {author} {\bibfnamefont
  {J.}~\bibnamefont {Reichel}}, \ and\ \bibinfo {author} {\bibfnamefont
  {J.~G.~E.}\ \bibnamefont {Harris}},\ }\href {\doibase
  http://dx.doi.org/10.1063/1.4768779} {\bibfield  {journal} {\bibinfo
  {journal} {Appl. Phys. Lett.}\ }\textbf {\bibinfo {volume} {101}},\ \bibinfo
  {eid} {221109} (\bibinfo {year} {2012})}\BibitemShut {NoStop}%
\bibitem [{\citenamefont {Sankey}\ \emph {et~al.}(2010)\citenamefont {Sankey},
  \citenamefont {Yang}, \citenamefont {Zwickl}, \citenamefont {Jayich},\ and\
  \citenamefont {Harris}}]{Sankey_tuning}%
  \BibitemOpen
  \bibfield  {author} {\bibinfo {author} {\bibfnamefont {J.~C.}\ \bibnamefont
  {Sankey}}, \bibinfo {author} {\bibfnamefont {C.}~\bibnamefont {Yang}},
  \bibinfo {author} {\bibfnamefont {B.~M.}\ \bibnamefont {Zwickl}}, \bibinfo
  {author} {\bibfnamefont {A.~M.}\ \bibnamefont {Jayich}}, \ and\ \bibinfo
  {author} {\bibfnamefont {J.~G.~E.}\ \bibnamefont {Harris}},\ }\href
  {http://dx.doi.org/10.1038/nphys1707} {\bibfield  {journal} {\bibinfo
  {journal} {Nature Phys.}\ }\textbf {\bibinfo {volume} {6}},\ \bibinfo {pages}
  {707–712} (\bibinfo {year} {2010})}\BibitemShut {NoStop}%
\bibitem [{\citenamefont {Santamore}\ \emph {et~al.}(2004)\citenamefont
  {Santamore}, \citenamefont {Doherty},\ and\ \citenamefont
  {Cross}}]{Santamore2004}%
  \BibitemOpen
  \bibfield  {author} {\bibinfo {author} {\bibfnamefont {D.~H.}\ \bibnamefont
  {Santamore}}, \bibinfo {author} {\bibfnamefont {A.~C.}\ \bibnamefont
  {Doherty}}, \ and\ \bibinfo {author} {\bibfnamefont {M.~C.}\ \bibnamefont
  {Cross}},\ }\href@noop {} {\bibfield  {journal} {\bibinfo  {journal} {Phys.\
  Rev.\ B}\ }\textbf {\bibinfo {volume} {70}},\ \bibinfo {pages} {144301}
  (\bibinfo {year} {2004})}\BibitemShut {NoStop}%
\bibitem [{\citenamefont {Jayich}\ \emph {et~al.}(2008)\citenamefont {Jayich},
  \citenamefont {Sankey}, \citenamefont {Zwickl}, \citenamefont {Yang},
  \citenamefont {Thompson}, \citenamefont {Girvin}, \citenamefont {Clerk},
  \citenamefont {Marquardt},\ and\ \citenamefont {Harris}}]{Jayich2008}%
  \BibitemOpen
  \bibfield  {author} {\bibinfo {author} {\bibfnamefont {A.~M.}\ \bibnamefont
  {Jayich}}, \bibinfo {author} {\bibfnamefont {J.~C.}\ \bibnamefont {Sankey}},
  \bibinfo {author} {\bibfnamefont {B.~M.}\ \bibnamefont {Zwickl}}, \bibinfo
  {author} {\bibfnamefont {C.}~\bibnamefont {Yang}}, \bibinfo {author}
  {\bibfnamefont {J.~D.}\ \bibnamefont {Thompson}}, \bibinfo {author}
  {\bibfnamefont {S.~M.}\ \bibnamefont {Girvin}}, \bibinfo {author}
  {\bibfnamefont {A.~A.}\ \bibnamefont {Clerk}}, \bibinfo {author}
  {\bibfnamefont {F.}~\bibnamefont {Marquardt}}, \ and\ \bibinfo {author}
  {\bibfnamefont {J.~G.~E.}\ \bibnamefont {Harris}},\ }\href@noop {} {\bibfield
   {journal} {\bibinfo  {journal} {New J. Phys.}\ }\textbf {\bibinfo {volume}
  {10}},\ \bibinfo {pages} {095008} (\bibinfo {year} {2008})}\BibitemShut
  {NoStop}%
\bibitem [{\citenamefont {Gangat}\ \emph {et~al.}(2011)\citenamefont {Gangat},
  \citenamefont {Stace},\ and\ \citenamefont {Milburn}}]{Gangat_Milburn}%
  \BibitemOpen
  \bibfield  {author} {\bibinfo {author} {\bibfnamefont {A.~A.}\ \bibnamefont
  {Gangat}}, \bibinfo {author} {\bibfnamefont {T.~M.}\ \bibnamefont {Stace}}, \
  and\ \bibinfo {author} {\bibfnamefont {G.~J.}\ \bibnamefont {Milburn}},\
  }\href@noop {} {\bibfield  {journal} {\bibinfo  {journal} {New J. Phys.}\
  }\textbf {\bibinfo {volume} {13}},\ \bibinfo {pages} {043024} (\bibinfo
  {year} {2011})}\BibitemShut {NoStop}%
\bibitem [{\citenamefont {Stannigel}\ \emph {et~al.}(2012)\citenamefont
  {Stannigel}, \citenamefont {Komar}, \citenamefont {Habraken}, \citenamefont
  {Bennett}, \citenamefont {Lukin}, \citenamefont {Zoller},\ and\ \citenamefont
  {Rabl}}]{Stannigel_Lukin}%
  \BibitemOpen
  \bibfield  {author} {\bibinfo {author} {\bibfnamefont {K.}~\bibnamefont
  {Stannigel}}, \bibinfo {author} {\bibfnamefont {P.}~\bibnamefont {Komar}},
  \bibinfo {author} {\bibfnamefont {S.~J.~M.}\ \bibnamefont {Habraken}},
  \bibinfo {author} {\bibfnamefont {S.~D.}\ \bibnamefont {Bennett}}, \bibinfo
  {author} {\bibfnamefont {M.~D.}\ \bibnamefont {Lukin}}, \bibinfo {author}
  {\bibfnamefont {P.}~\bibnamefont {Zoller}}, \ and\ \bibinfo {author}
  {\bibfnamefont {P.}~\bibnamefont {Rabl}},\ }\href {\doibase
  10.1103/PhysRevLett.109.013603} {\bibfield  {journal} {\bibinfo  {journal}
  {Phys. Rev. Lett.}\ }\textbf {\bibinfo {volume} {109}},\ \bibinfo {pages}
  {013603} (\bibinfo {year} {2012})}\BibitemShut {NoStop}%
\bibitem [{\citenamefont {Ludwig}(2013)}]{Ludwig_thesis2013}%
  \BibitemOpen
  \bibfield  {author} {\bibinfo {author} {\bibfnamefont {M.}~\bibnamefont
  {Ludwig}},\ }\emph {\bibinfo {title} {{Collective quantum effects in
  optomechanical systems}}},\ \href@noop {} {Ph.D. thesis},\ \bibinfo  {school}
  {University of Erlangen-Nuremberg} (\bibinfo {year} {2013})\BibitemShut
  {NoStop}%
\bibitem [{\citenamefont {Kalaee}\ \emph {et~al.}(2015)\citenamefont {Kalaee},
  \citenamefont {Para\"iso},\ and\ \citenamefont {Painter}}]{Kalaee2015}%
  \BibitemOpen
  \bibfield  {author} {\bibinfo {author} {\bibfnamefont {M.}~\bibnamefont
  {Kalaee}}, \bibinfo {author} {\bibfnamefont {T.~K.}\ \bibnamefont
  {Para\"iso}}, \ and\ \bibinfo {author} {\bibfnamefont {O.}~\bibnamefont
  {Painter}},\ }\href@noop {} {\enquote {\bibinfo {title} {Design of a quasi-2d
  photonic crystal optomechanical cavity with tunable, large $x^2$-coupling},}\
  } (\bibinfo {year} {2015})\BibitemShut {NoStop}%
\bibitem [{\citenamefont {Johnson}\ and\ \citenamefont
  {Joannopoulos}(2001)}]{Johnson2001}%
  \BibitemOpen
  \bibfield  {author} {\bibinfo {author} {\bibfnamefont {S.~G.}\ \bibnamefont
  {Johnson}}\ and\ \bibinfo {author} {\bibfnamefont {J.~D.}\ \bibnamefont
  {Joannopoulos}},\ }\href
  {http://www.opticsexpress.org/abstract.cfm?URI=OPEX-8-3-173} {\bibfield
  {journal} {\bibinfo  {journal} {Opt. Express}\ }\textbf {\bibinfo {volume}
  {8}},\ \bibinfo {pages} {173} (\bibinfo {year} {2001})}\BibitemShut {NoStop}%
\bibitem [{MPB()}]{MPB}%
  \BibitemOpen
  \href@noop {} {}\bibinfo {note} {The MIT Photonic-Bands (MPB) software
  package, http://ab-initio.mit.edu/mpb}\BibitemShut {NoStop}%
\bibitem [{COM()}]{COMSOL}%
  \BibitemOpen
  \href@noop {} {}\bibinfo {note} {COMSOL Multiphysics 3.5,
  http://www.comsol.com/}\BibitemShut {NoStop}%
\bibitem [{\citenamefont {Safavi-Naeini}\ \emph {et~al.}(2010)\citenamefont
  {Safavi-Naeini}, \citenamefont {Alegre}, \citenamefont {Winger},\ and\
  \citenamefont {Painter}}]{Safavi-Naeini_APL_2010}%
  \BibitemOpen
  \bibfield  {author} {\bibinfo {author} {\bibfnamefont {A.~H.}\ \bibnamefont
  {Safavi-Naeini}}, \bibinfo {author} {\bibfnamefont {T.~P.~M.}\ \bibnamefont
  {Alegre}}, \bibinfo {author} {\bibfnamefont {M.}~\bibnamefont {Winger}}, \
  and\ \bibinfo {author} {\bibfnamefont {O.}~\bibnamefont {Painter}},\ }\href
  {\doibase http://dx.doi.org/10.1063/1.3507288} {\bibfield  {journal}
  {\bibinfo  {journal} {Appl. Phys. Lett.}\ }\textbf {\bibinfo {volume} {97}},\
  \bibinfo {eid} {181106} (\bibinfo {year} {2010})}\BibitemShut {NoStop}%
\bibitem [{\citenamefont {Winger}\ \emph {et~al.}(2011)\citenamefont {Winger},
  \citenamefont {Blasius}, \citenamefont {Alegre}, \citenamefont
  {Safavi-Naeini}, \citenamefont {Meenehan}, \citenamefont {Cohen},
  \citenamefont {Stobbe},\ and\ \citenamefont {Painter}}]{Winger:chip-scale}%
  \BibitemOpen
  \bibfield  {author} {\bibinfo {author} {\bibfnamefont {M.}~\bibnamefont
  {Winger}}, \bibinfo {author} {\bibfnamefont {T.~D.}\ \bibnamefont {Blasius}},
  \bibinfo {author} {\bibfnamefont {T.~P.~M.}\ \bibnamefont {Alegre}}, \bibinfo
  {author} {\bibfnamefont {A.~H.}\ \bibnamefont {Safavi-Naeini}}, \bibinfo
  {author} {\bibfnamefont {S.}~\bibnamefont {Meenehan}}, \bibinfo {author}
  {\bibfnamefont {J.}~\bibnamefont {Cohen}}, \bibinfo {author} {\bibfnamefont
  {S.}~\bibnamefont {Stobbe}}, \ and\ \bibinfo {author} {\bibfnamefont
  {O.}~\bibnamefont {Painter}},\ }\href {\doibase 10.1364/OE.19.024905}
  {\bibfield  {journal} {\bibinfo  {journal} {Opt. Express}\ }\textbf {\bibinfo
  {volume} {19}},\ \bibinfo {pages} {24905} (\bibinfo {year}
  {2011})}\BibitemShut {NoStop}%
\bibitem [{\citenamefont {Michael}\ \emph {et~al.}(2007)\citenamefont
  {Michael}, \citenamefont {Borselli}, \citenamefont {Johnson}, \citenamefont
  {Chrystal},\ and\ \citenamefont {Painter}}]{Michael:07}%
  \BibitemOpen
  \bibfield  {author} {\bibinfo {author} {\bibfnamefont {C.~P.}\ \bibnamefont
  {Michael}}, \bibinfo {author} {\bibfnamefont {M.}~\bibnamefont {Borselli}},
  \bibinfo {author} {\bibfnamefont {T.~J.}\ \bibnamefont {Johnson}}, \bibinfo
  {author} {\bibfnamefont {C.}~\bibnamefont {Chrystal}}, \ and\ \bibinfo
  {author} {\bibfnamefont {O.}~\bibnamefont {Painter}},\ }\href {\doibase
  10.1364/OE.15.004745} {\bibfield  {journal} {\bibinfo  {journal} {Opt.
  Express}\ }\textbf {\bibinfo {volume} {15}},\ \bibinfo {pages} {4745}
  (\bibinfo {year} {2007})}\BibitemShut {NoStop}%
\bibitem [{\citenamefont {Eichenfield}\ \emph
  {et~al.}(2009{\natexlab{b}})\citenamefont {Eichenfield}, \citenamefont
  {Camacho}, \citenamefont {Chan}, \citenamefont {Vahala},\ and\ \citenamefont
  {Painter}}]{Eichenfield:09}%
  \BibitemOpen
  \bibfield  {author} {\bibinfo {author} {\bibfnamefont {M.}~\bibnamefont
  {Eichenfield}}, \bibinfo {author} {\bibfnamefont {R.}~\bibnamefont
  {Camacho}}, \bibinfo {author} {\bibfnamefont {J.}~\bibnamefont {Chan}},
  \bibinfo {author} {\bibfnamefont {K.~J.}\ \bibnamefont {Vahala}}, \ and\
  \bibinfo {author} {\bibfnamefont {O.}~\bibnamefont {Painter}},\ }\href
  {http://dx.doi.org/10.1038/nature08061} {\bibfield  {journal} {\bibinfo
  {journal} {Nature}\ }\textbf {\bibinfo {volume} {459}},\ \bibinfo {pages}
  {550} (\bibinfo {year} {2009}{\natexlab{b}})}\BibitemShut {NoStop}%
\bibitem [{\citenamefont {Bao}\ and\ \citenamefont {Yang}(2007)}]{Bao2007}%
  \BibitemOpen
  \bibfield  {author} {\bibinfo {author} {\bibfnamefont {M.}~\bibnamefont
  {Bao}}\ and\ \bibinfo {author} {\bibfnamefont {H.}~\bibnamefont {Yang}},\
  }\href@noop {} {\bibfield  {journal} {\bibinfo  {journal} {Sensors and
  Actuators A-Physical}\ }\textbf {\bibinfo {volume} {136}},\ \bibinfo {pages}
  {3} (\bibinfo {year} {2007})}\BibitemShut {NoStop}%
\bibitem [{\citenamefont {Lee}\ \emph {et~al.}(2015)\citenamefont {Lee},
  \citenamefont {Underwood}, \citenamefont {Mason}, \citenamefont {Shkarin},
  \citenamefont {Hoch},\ and\ \citenamefont {Harris}}]{Lee2015-x2spring}%
  \BibitemOpen
  \bibfield  {author} {\bibinfo {author} {\bibfnamefont {D.}~\bibnamefont
  {Lee}}, \bibinfo {author} {\bibfnamefont {M.}~\bibnamefont {Underwood}},
  \bibinfo {author} {\bibfnamefont {D.}~\bibnamefont {Mason}}, \bibinfo
  {author} {\bibfnamefont {A.~B.}\ \bibnamefont {Shkarin}}, \bibinfo {author}
  {\bibfnamefont {S.~W.}\ \bibnamefont {Hoch}}, \ and\ \bibinfo {author}
  {\bibfnamefont {J.~G.~E.}\ \bibnamefont {Harris}},\ }\href {\doibase
  10.1038/ncomms7232} {\bibfield  {journal} {\bibinfo  {journal} {Nat.
  Commun.}\ }\textbf {\bibinfo {volume} {6}},\ \bibinfo {pages} {6232}
  (\bibinfo {year} {2015})}\BibitemShut {NoStop}%
\bibitem [{\citenamefont {Sekoguchi}\ \emph {et~al.}(2014)\citenamefont
  {Sekoguchi}, \citenamefont {Takahashi}, \citenamefont {Asano},\ and\
  \citenamefont {Noda}}]{Sekoguchi2014-9millionQ}%
  \BibitemOpen
  \bibfield  {author} {\bibinfo {author} {\bibfnamefont {H.}~\bibnamefont
  {Sekoguchi}}, \bibinfo {author} {\bibfnamefont {Y.}~\bibnamefont
  {Takahashi}}, \bibinfo {author} {\bibfnamefont {T.}~\bibnamefont {Asano}}, \
  and\ \bibinfo {author} {\bibfnamefont {S.}~\bibnamefont {Noda}},\ }\href
  {\doibase 10.1364/OE.22.000916} {\bibfield  {journal} {\bibinfo  {journal}
  {Opt. Express}\ }\textbf {\bibinfo {volume} {22}},\ \bibinfo {pages} {916}
  (\bibinfo {year} {2014})}\BibitemShut {NoStop}%
\bibitem [{\citenamefont {Meenehan}\ \emph {et~al.}(2015)\citenamefont
  {Meenehan}, \citenamefont {Cohen}, \citenamefont {MacCabe}, \citenamefont
  {Marsili}, \citenamefont {Shaw},\ and\ \citenamefont
  {Painter}}]{Meenehan2014}%
  \BibitemOpen
  \bibfield  {author} {\bibinfo {author} {\bibfnamefont {S.~M.}\ \bibnamefont
  {Meenehan}}, \bibinfo {author} {\bibfnamefont {J.~D.}\ \bibnamefont {Cohen}},
  \bibinfo {author} {\bibfnamefont {G.~S.}\ \bibnamefont {MacCabe}}, \bibinfo
  {author} {\bibfnamefont {F.}~\bibnamefont {Marsili}}, \bibinfo {author}
  {\bibfnamefont {M.~D.}\ \bibnamefont {Shaw}}, \ and\ \bibinfo {author}
  {\bibfnamefont {O.}~\bibnamefont {Painter}},\ }\href@noop {} {\bibfield
  {journal} {\bibinfo  {journal} {arXiv:1503.05135}\ } (\bibinfo {year}
  {2015})}\BibitemShut {NoStop}%
\bibitem [{\citenamefont {Chan}\ \emph {et~al.}(2012)\citenamefont {Chan},
  \citenamefont {Safavi-Naeini}, \citenamefont {Hill}, \citenamefont
  {Meenehan},\ and\ \citenamefont {Painter}}]{Chan2012}%
  \BibitemOpen
  \bibfield  {author} {\bibinfo {author} {\bibfnamefont {J.}~\bibnamefont
  {Chan}}, \bibinfo {author} {\bibfnamefont {A.~H.}\ \bibnamefont
  {Safavi-Naeini}}, \bibinfo {author} {\bibfnamefont {J.~T.}\ \bibnamefont
  {Hill}}, \bibinfo {author} {\bibfnamefont {S.}~\bibnamefont {Meenehan}}, \
  and\ \bibinfo {author} {\bibfnamefont {O.}~\bibnamefont {Painter}},\
  }\href@noop {} {\bibfield  {journal} {\bibinfo  {journal} {Appl.\ Phys.\
  Lett.}\ }\textbf {\bibinfo {volume} {101}},\ \bibinfo {pages} {081115}
  (\bibinfo {year} {2012})}\BibitemShut {NoStop}%
\bibitem [{\citenamefont {Pitanti}\ \emph {et~al.}(2015)\citenamefont
  {Pitanti}, \citenamefont {Fink}, \citenamefont {Safavi-Naeini}, \citenamefont
  {Hill}, \citenamefont {Lei}, \citenamefont {Tredicucci},\ and\ \citenamefont
  {Painter}}]{Pitanti2015}%
  \BibitemOpen
  \bibfield  {author} {\bibinfo {author} {\bibfnamefont {A.}~\bibnamefont
  {Pitanti}}, \bibinfo {author} {\bibfnamefont {J.~M.}\ \bibnamefont {Fink}},
  \bibinfo {author} {\bibfnamefont {A.~H.}\ \bibnamefont {Safavi-Naeini}},
  \bibinfo {author} {\bibfnamefont {J.~T.}\ \bibnamefont {Hill}}, \bibinfo
  {author} {\bibfnamefont {C.~U.}\ \bibnamefont {Lei}}, \bibinfo {author}
  {\bibfnamefont {A.}~\bibnamefont {Tredicucci}}, \ and\ \bibinfo {author}
  {\bibfnamefont {O.}~\bibnamefont {Painter}},\ }\href@noop {} {\bibfield
  {journal} {\bibinfo  {journal} {Opt. Express}\ }\textbf {\bibinfo {volume}
  {23}},\ \bibinfo {pages} {3196} (\bibinfo {year} {2015})}\BibitemShut
  {NoStop}%
\end{thebibliography}
%

\end{document}